\documentclass[article,preprint,prd,aps,showkeys,nofootinbib]{revtex4}
\usepackage{graphics}
\usepackage{cancel}
\usepackage{ulem}
\usepackage{color}
\usepackage{CJK}
\usepackage{graphicx}
\usepackage{dcolumn}
\usepackage{bm}
\usepackage[colorlinks=true,linkcolor=blue,citecolor=blue, urlcolor=blue]{hyperref}

\begin{document}

\title{Neutralino dark matter in the extension of MSSM with two triplets and singlet}

\author{
Zhong-Jun Yang$^{1}$\footnote{zj\_yang@cqu.edu.cn}
, Jin-Lei Yang$^{2}$\footnote{jlyang@hbu.edu.cn}
, Shu-Min Zhao$^{2}$\footnote{zhaosm@hbu.edu.cn}
, Xing-Gang Wu$^{1}$\footnote{wuxg@cqu.edu.cn}
, Tai-Fu Feng$^{1,2}$\footnote{fengtf@hbu.edu.cn}
}

\affiliation{$^1$ Department of Physics, Chongqing Key Laboratory for Strongly Coupled Physics, Chongqing University, Chongqing 401331, P.R. China \\
$^2$ Department of Physics, Key Laboratory of High-precision Computation and Application of Quantum Field Theory of Hebei Province, Hebei University, Baoding 071002, P.R. China}

\date{\today}

\begin{abstract}

In an extension of MSSM with two triplets and a singlet, called the TNMSSM, there are seven neutralinos which can enrich the study of cold dark matter if one expects that the weakly interacting massive particle (WIMP) is responsible for the observation of Planck satellite. Such a model, compared to the MSSM, can naturally offer a solution to the $\mu$ problem, and its lightest neutralino, which is bino-like, can also provide a correct relic density by using the coannihilation mechanism due to the newly added triplinos. Taking into account the related experimental measurements, such as the bound on the SM-like Higgs mass, the $B$ meson rare decays, the anomalous magnetic moment of the muon $a_\mu$, the Large Hadron Collider (LHC) measurements and the latest dark matter direct detection experiment LUX-ZEPLIN (LZ), the TNMSSM parameter space can be strictly limited. In respect to all the constraints mentioned above, we find that a bino-like neutralino with a mass in the region $[100, 450]~\rm{GeV}$ can successfully account for the correct dark matter relic density. Additionally, most of the viable parameter space can be tested in the near future experiments such as the Xenon-nT experiment or LHC.

\end{abstract}


\maketitle
\newpage
\tableofcontents
\newpage

\section{Introduction}

Even though at the present all the LHC measurements are still in agreement with the Standard Model (SM) predictions within reasonable errors, due to several shortcomings of the SM, theoretical physicists have been actively exploring the new physics (NP) beyond SM. The Minimal Supersymmetric Standard Model (MSSM)~\cite{Haber:1984rc}, as a famous extension of the SM in Supersymmetry (SUSY), can offer a solution to the notorious hierarchy problem inherent in the SM and provide a framework for the unification of gauge coupling constants~\cite{Cao:2011sn}. Additionally, in the MSSM, there are also many dark matter candidates that can explain the relic density $\Omega_{\rm{DM}} h^2=0.120\pm0.001$ observed by the Planck satellite~\cite{Planck}. The lightest supersymmetric particle (LSP), for instance, often referred to as a weakly interacting massive particle (WIMP), can exhibit a remarkably suitable annihilation cross section for the cold dark matter freeze-out production mechanism. This is known as the ``WIMP miracle''. Without the help of WIMPs, of course, other rational dark matter candidates (e.g. axions~\cite{PQM,Weinberg:1977ma,landscape}, sterile neutrinos~\cite{Dodelson:1993je}, Primordial Black Holes (PBHs)~\cite{Carr:2016drx}, etc.) can also appear in some specific NP models and can give reasonable explanations for the Planck observation. Despite the numerous advantages of the MSSM over the SM, there are still many issues that are not well understood, such as the well-known SUSY $\mu$ problem. The higgsino mass parameter $\mu$ is introduced via the superpotential term $\mu H_u H_d$, where $H_u$ and $H_d$ represent the two Higgs doublet superfields. Theoretically, the natural value for this SUSY-conserving $\mu$ parameter would appear to be at the order of the Planck scale. However, in order to maintain the consistency with the physical phenomenology\footnote{The naturalness arguments, e.g., which mainly focus on the fine-tuning of parameters in the MSSM Higgs potential minimization condition $\frac{M_Z^2}{2}=\frac{m_{H_d}^2-m_{H_u}^2\tan^2\beta}{\tan^2\beta-1}-\mu^2$, suggest that $\mu$ should be at the weak scale ($|\mu| \sim 100 - 300 \rm{GeV}$ is claimed in Ref.~\cite{Bae:2019dgg} and $\mu\sim 100 - 350 ~\rm{GeV}$ is claimed in Ref.~\cite{Baer:2018avn}). Intuitively, if $\mu\gg M_Z$, the first term on the right side of this equation must also be large to achieve a significant cancellation with $\mu^2$ in order to reproduce $M_Z$. In this sense, one can have a suspicion that the models with $\mu\gg M_Z$ might be subject to fine-tuning.}, $\mu$ needs to be constrained to the weak scale. While a moderately larger $\mu$ (e.g., multi-$\rm{TeV}$) can be tolerated, such a $\mu$ is still far too small compared to the Planck scale.

Although the MSSM neutralino dark matter still survives, the dark matter direct detection experiments have ruled out a large amount of MSSM parameter space that was once considered promising. The lightest MSSM neutralino can be classified as bino-like, wino-like, higgsino-like or other well-tempered types, determined by the parameters $M_1, M_2, \mu$ and $\tan\beta$. The relic density can be determined by calculating the cross section of the annihilation or coannihilation processes involving the next-to-lightest supersymmetric particle (NLSP). If the LSP is wino-like, considering the ``Sommerfeld enhancement" effect~\cite{SOMMERFELD}, it can constitute the entire dark matter content only if its mass is approximately $2.9~\rm{TeV}$. In the case of a pure higgsino, the parameter $\mu$ would need to be around $1.1~\rm{TeV}$. Both cases make the theory unappealing\footnote{In such two cases, the LSP with a multi-$\rm{TeV}$ mass means a heavier spectrum of SUSY particles, and this would lead to many concerns. Theoretically, a heavier SUSY particle spectrum weakens many quantum correction effects, reducing the model's ability to explain phenomena beyond the SM to some extent. Additionally, to achieve proper electroweak symmetry breaking and maintain the correct Higgs mass, several parameters in the SUSY model, such as scalar mass-squared soft-breaking terms and trilinear soft-breaking parameters, also require more precise fine-tuning. Experimentally, given the current energy and luminosity of the LHC, it is quite a challenging task to effectively produce and detect a SUSY particle spectrum with such heavy masses. These factors significantly reduce the appeal of the SUSY model with such configurations.}. Conversely, if the LSP is bino-like, it results in an excessively large relic density, which cannot account for the Planck observation, and some coannihilation processes (involving the next-lightest neutralino, lightest chargino or sleptons) are needed.

There are also many NP models (e.g. B-LSSM~\cite{FileviezPerez:2008sx, Ambroso:2010pe, Yang:2021duj}, BLMSSM~\cite{FileviezPerez:2010gw, Zhao:2016jcx, Zhao:2018won}, $\mu\nu$SSM~\cite{Lopez-Fogliani:2005vcg, Escudero:2008jg, Liu:2020mev}, $\rm{U(1)_X}$SSM~\cite{Staub:2013tta, Zhao:2021eaa}, NMSSM~\cite{Ellwanger:2009dp, Cao:2011sn}, TMSSM~\cite{Espinosa:1991gr, Espinosa:1992hp, Espinosa:1998re}, TNMSSM~\cite{Agashe:2011ia}, etc.) that are very promising other than the MSSM. For example, the TNMSSM, being an extension of MSSM with two $\rm{SU(2)_L}$ triplets and one singlet, can naturally tackle the $\mu$ problem. Furthermore, this model can also relieve the so-called little hierarchy problem, which is manifested as the measured mass of the SM-like Higgs boson being much heavier than its tree-level theoretical mass. For $\mu$ problem, the superpotential has a $\rm{Z_3}$ symmetry and, therefore, a bare $\mu$ term is forbidden. The effective $\mu$ term is dynamically generated by a vacuum expectation value (VEV) of the singlet just like the NMSSM case. For the little hierarchy problem, the triplets can improve the SM-like Higgs mass at tree level, and then large loop corrections to the $125~\rm{GeV}$ Higgs mass are not necessary. Besides, small Majorana masses for the neutrinos can also be easily obtained if adding a term like $\xi \hat{l} \cdot \hat{T} \hat{l}$ with very small dimensionless Yukawa coupling $\xi$ to the potential, where $\hat{l}$ is the lepton $\rm{SU(2)_L}$ doublet superfield and $\hat{T}$ is the triplet superfield with hypercharge $+1$.

Moreover, in the TNMSSM, there are seven neutralinos due to the presence of the additional singlino $\tilde{S}$ and triplinos $\tilde{T}$ and $\tilde{\bar{T}}$. The composition of the dark matter particle in the TNMSSM, if one expects the lightest neutralino being the LSP, can have more possibilities compared to the MSSM scenarios mentioned above. Given the increased number of neutralinos and the various mixing possibilities, studying the lightest neutralino as the dark matter candidate in the TNMSSM becomes particularly intriguing. Therefore, we mainly focus on this in this work. In addition, when we consider the latest experimental data of anomalous magnetic moment of the muon $a_\mu:=(g-2)_\mu/2$ (the Brookhaven National Laboratory (BNL) E821 measurement \cite{Muong-2:2006rrc} and the Fermilab National Accelerator Laboratory Muon Experiment (FNAL) \cite{g-2 SM EXP}), more constraints should be imposed on the parameter space. The SM prediction of muon anomaly is $a^{\rm{sm}}_\mu = 116591810 (43) \times 10^{-11} (0.37 \rm{ppm})$~\cite{g-2 SM 1, g-2 SM 2, g-2 SM 3}, and the latest averaged experimental value of muon anomaly is $a^{\rm{exp}}_\mu = 116592059 (22) \times 10^{-11} (0.19 \rm{ppm})$~\cite{g-2 SM EXP}. The deviation between the experiment and the SM prediction then is $\Delta a_\mu^{\rm{exp}} = a^{\rm{exp}}_\mu - a^{\rm{sm}}_\mu= 249 (48) \times 10^{-11}$, which is $5.1\sigma$. This anomaly is an important hint of NP phenomenology, and therefore we will take this into account in our following analysis. It is also worth noting that the lattice calculations, carried out by Budapest-Marseille-Wuppertal (BMW), for the leading-order hadronic vacuum polarization (LO-HVP) contribution to the muon anomalous
magnetic moment~\cite{Kotov:2023wug,Toth:2022lsa,BMW:2022rdk,Budapest-Marseille-Wuppertal:2017sdk,Budapest-Marseille-Wuppertal:2018ivi} indicates the corrections to $a_\mu$ from NP models are not necessary. Given that this result still awaits independent confirmation, we consequently do not take this result into account in this work.

The remaining parts of the paper are organized as follows. In Sec.~\ref{Sec II}, we provide an introduction to the TNMSSM. In Sec.~\ref{Sec III}, we give the method of calculation about the muon anomaly $a_\mu$, the dark matter relic density involving the coannihilation and the WIMP-nucleon scattering cross section of direct detection. Numerical results and discussions are given in Sec.~\ref{Sec IV}, and Sec.~\ref{Sec V} is reserved as a summary. In the Appendix, we give the tadpole equations of the TNMSSM and the Feynman diagrams involving coannihilation.

\section{Framework of the model}\label{Sec II}
In the TNMSSM, the superpotential, which possesses a $\rm{Z_3}$ symmetry, is given by
\begin{eqnarray}
W&=&\hat{s}\left(\lambda \hat{H}_{u} \cdot \hat{H}_{d}+\lambda_{T} {tr} (\hat{\bar{T}} \hat{T})\right)+\frac{\kappa}{3} \hat{s}^{3}+\chi_{u} \hat{H}_{u} \cdot \hat{\bar{T}} \hat{H}_{u}+\chi_{d} \hat{H}_{d} \cdot \hat{T} \hat{H}_{d}\nonumber\\
&&+y_{u} \hat{H}_{u} \cdot \hat{q} \hat{u}-y_{d} \hat{H}_{d} \cdot \hat{q} \hat{d}-y_{e} \hat{H}_{d} \cdot \hat{l} \hat{e},\label{spW}
\end{eqnarray}
where $\lambda, \lambda_T, \kappa, \chi_u, \chi_d$ are dimensionless couplings and $y_u, y_d, y_e$ are Yukawa couplings as usual. In general, there are many sources of the CP violation in this model. For instance, an additional physical CP-violating phase, Arg$(\chi_u\chi_d\kappa\lambda_T^*(\lambda^*)^{2})$, arise from the superpotential compared to the MSSM case. In this work, for the sake of simplicity, we only restrict ourself to the CP-conserving case. The information about the chiral superfields in Eq.~(\ref{spW}) are given in Table \ref{tab:charges}. $\hat{s}$ is the gauge singlet superfield, and the $\rm{SU(2)_L}$ triplet superfields with hypercharge $ \pm 1 $ are defined as follows:
\begin{eqnarray}
T \equiv T^{a} \sigma^{a}=\left(\begin{array}{cc}
T^{+} / \sqrt{2} & -T^{++} \\
T^{0} & -T^{+} / \sqrt{2}
\end{array}\right),~
\bar{T} \equiv \bar{T}^{a} \sigma^{a}=\left(\begin{array}{cc}
\bar{T}^{-} / \sqrt{2} & -\bar{T}^{0} \\
\bar{T}^{--} & -\bar{T}^{-} / \sqrt{2}
\end{array}\right).
\end{eqnarray}
where $\sigma^{a}$ are $ 2 \times 2 $ Pauli matrices. Using these definitions, the products in the first line of superpotential $W$ can be written as follows:
\begin{eqnarray}
H_{u} \cdot H_{d} &=& H_{u}^{+} H_{d}^{-}-H_{u}^{0} H_{d}^{0}, \nonumber\\
H_{u} \cdot \bar{T} H_{u} &=&\left(H_{u}^{0}\right)^{2} \bar{T}^{0}+\left(H_{u}^{+}\right)^{2} \bar{T}^{--}- \sqrt{2} H_{u}^{+} H_{u}^{0} \bar{T}^{-}, \nonumber\\
H_{d} \cdot T H_{d} &=&\left(H_{d}^{0}\right)^{2} T^{0}+\left(H_{d}^{-}\right)^{2} T^{++}- \sqrt{2} H_{d}^{-} H_{d}^{0} T^{+}.
\end{eqnarray}
\begin{table}[htb]
\begin{center}
\begin{minipage}[]{0.95\linewidth}\caption{The information about the gauge charges of the chiral superfields in the TNMSSM. \label{tab:charges}}
\end{minipage}
\vspace{0.3cm}
\begin{tabular}{|c|c|c|c|c|}
\hline
\text { SuperField } & \text { Spin } 0 & \text { Spin } $\frac{1}{2}$ & \text{ Generations } & $\rm{U(1)_Y \otimes SU(2)_L \otimes SU(3)_C}$ \\
\hline
\hline
 $\hat{q}$ & $\tilde{q}$ & $q$ & 3 & $\left(\frac{1}{6}, \mathbf{2}, \mathbf{3}\right)$ \\
$\hat{l}$ & $\tilde{l}$ & $l$ & 3 & $\left(-\frac{1}{2}, \mathbf{2}, \mathbf{1}\right)$ \\
$\hat{H}_{d}$ & $H_{d}$ & $\tilde{H}_{d}$ & 1 & $\left(-\frac{1}{2}, \mathbf{2}, \mathbf{1}\right)$ \\
$\hat{H}_{u}$ & $H_{u}$ & $\tilde{H}_{u}$ & 1 & $\left(\frac{1}{2}, \mathbf{2}, \mathbf{1}\right)$ \\
$\hat{d}$ & $\tilde{d}_{R}^{*}$ & $d_{R}^{*}$ & 3 & $\left(\frac{1}{3}, \mathbf{1}, \overline{\mathbf{3}}\right)$ \\
$\hat{u}$ & $\tilde{u}_{R}^{*}$ & $u_{R}^{*}$ & 3 & $\left(-\frac{2}{3}, \mathbf{1}, \overline{\mathbf{3}}\right)$ \\
$\hat{e}$ & $\tilde{e}_{R}^{*}$ & $e_{R}^{*}$ & 3 & $(1, \mathbf{1}, \mathbf{1})$ \\
$\hat{s}$ & $S$ & $\tilde{S}$ & 1 & $(0, \mathbf{1}, \mathbf{1})$ \\
$\hat{T}$ & $T$ & $\tilde{T}$ & 1 & $(1, \mathbf{3}, \mathbf{1})$ \\
$\hat{\bar{T}}$ & $\bar{T}$ & $\tilde{\bar{T}}$ & 1 & $(-1, \mathbf{3}, \mathbf{1})$ \\
\hline
\end{tabular}
\end{center}
\end{table}
Besides, the soft breaking terms in the TNMSSM are given by
\begin{eqnarray}
-\mathcal{L}_{\mathrm{soft}} & = &+  m_{H_{u}}^{2}\left|H_{u}\right|^{2}+m_{H_{d}}^{2}\left|H_{d}\right|^{2}+m_{S}^{2}|S|^{2}+m_{T}^{2} {tr}\left(|T|^{2}\right)+m_{\bar{T}}^{2} {tr}\left(|\bar{T}|^{2}\right) \nonumber\\
&&+\tilde{u}_{L,i\alpha}^{*}\delta_{\alpha\beta}m_{\tilde{q},i j}^{2}\tilde{u}_{L,j\beta}+\tilde{d}_{L,i\alpha}^{*}\delta_{\alpha\beta}m_{\tilde{q},ij}^{2}\tilde{d}_{L,j\beta} +\tilde{\nu}_{L,i}^{*}m_{\tilde{l},i j}^{2}\tilde{\nu}_{L,j}+\tilde{e}_{L,i}^{*}m_{\tilde{l},i j}^{2}\tilde{e}_{L,j}\nonumber\\
&&+\tilde{u}_{R,i\alpha}^{*}\delta_{\alpha\beta}m_{\tilde{u},i j}^{2}\tilde{u}_{R,j\beta}+\tilde{d}_{R,i\alpha}^{*}\delta_{\alpha\beta}m_{\tilde{d},ij}^{2}\tilde{d}_{R,j\beta}+\tilde{e}_{R,i}^{*}m_{\tilde{e},i j}^{2}\tilde{e}_{R,j}\nonumber\\
&&+ (S(T_{\lambda} H_{u} \cdot H_{d}+T_{\lambda_T} {tr}(T \bar{T}))+\frac{T_{\kappa}}{3} S^{3}+T_{\chi_u} H_{u} \cdot \bar{T} H_{u}+T_{\chi_d} H_{d} \cdot T H_{d} \nonumber\\
&&-H_u^+\tilde{u}_{R,i\alpha}^*\delta_{\alpha\beta}\tilde{d}_{L,j\beta}T_{u,i j}+H_u^0\tilde{u}_{R,i\alpha}^*\delta_{\alpha\beta}\tilde{u}_{L,j\beta}T_{u,i j}+H^0_{d}\tilde{d}^{*}_{R,i\alpha}\delta_{\alpha\beta}\tilde{d}_{L,j\beta}T_{d,ij}\nonumber\\
&&-H^-_{d}\tilde{d}^*_{R,i\alpha}\delta_{\alpha\beta}\tilde{u}_{L,j\beta}T_{d,i j}+H^0_d\tilde{e}^*_{R,i}\tilde{e}_{L,j}T_{e,i j}-H^-_{d}\tilde{e}^*_{R,i}\tilde{\nu}_{L,j}T_{e,i j}+\text {h.c.})\nonumber\\
&&+ \frac{1}{2}(M_1 {\tilde{B}^{0}}{\tilde{B}^{0}}+M_2 \delta_{ij}{\tilde{W}^i}{\tilde{W}^j}+M_3 \delta_{\alpha\beta}{\tilde{g}^\alpha}{\tilde{g}^\beta} +\text {h.c.}),
\end{eqnarray}
where h.c. represents the terms of Hermitian conjugation. As can be seen from above equations, the SUSY $\mu$ and $B \mu$ terms (so do $\mu_T$ and $B \mu_T$ terms) are firstly absent due to the imposed $\rm{Z_{3}}$ symmetry, and then regenerated once the scalar $S$ receives a VEV. In this model, the scalars that acquire VEVs are parameterized as:
\begin{eqnarray}
H_{d}^{0} & = & \frac{1}{\sqrt{2}}( \phi_{d}+ v_{d}+i  \sigma_{d}),~H_{u}^{0}  =  \frac{1}{\sqrt{2}}( \phi_{u}+ v_{u}+i  \sigma_{u}),~S  =  \frac{1}{\sqrt{2}}( \phi_{s}+ v_{s}+i  \sigma_{s}), \nonumber\\
T^{0} & = & \frac{1}{\sqrt{2}}( \phi_{T}+ v_{T}+i  \sigma_{T}),~\bar{T}^{0}  =  \frac{1}{\sqrt{2}}( \phi_{\bar{T}}+ v_{\bar{T}}+i  \sigma_{\bar{T}}).
\end{eqnarray}
Along with the scalars obtaining VEVs, not only the SUSY $\mu=\frac{1}{\sqrt{2}} \lambda v_s$ and $B\mu=\frac{1}{\sqrt{2}} T_{\lambda} v_s$ regenerate, but also the local gauge group $\rm{SU(2)_L\otimes U(1)_Y}$ breaks down to the electromagnetic symmetry $\rm{U(1)_{em}}$. The scalar potential and tadpole equations are given in the Appendix~\ref{APPENDIX A}.

In order to convey more information of this model to readers, we give some mass matrice that quite different from them in MSSM. The tree-level squared mass of $W$ boson, in the TNMSSM, is $m_W^2=\frac{1}{4} g_2^2 \left(2 v_{T\bar{T}}^2+v_{ud}^2\right)$, and that of $Z$ boson is $m_Z^2=\frac{1}{4} \left(g_1^2+g_2^2\right) \left(4 v_{T\bar{T}}^2+v_{ud}^2\right)$. Here $v_{T\bar{T}}^2=v_{T}^2+v_{\bar{T}}^2$ and $v_{ud}^2=v_{d}^2+v_{u}^2$. For convenience, we define $\tan \beta^{\prime}=v_{T}/v_{\bar{T}}$, and this is analogous to the definition of $\tan \beta=v_{u}/v_{d}$ in MSSM and also in this model. For CP-even Higgses, they are made up of $\phi_d,\phi_u,\phi_s,\phi_T$ and $\phi_{\bar{T}}$, and the mass matrix is $5\times5$. Since the matrix elements are too lengthy, we do not show them here. For the SM-like Higgs, we consider its mass up to two-loop level, $m_h = (m_h^2|_{\rm{tree}} +m_h^2|_{\rm{one-loop}} + m_h^2|_{\rm{two-loop}})^{\frac{1}{2}}$, and the analytical calculation for the loop corrected mass can be found in many literature, e.g., Refs.~\cite{Carena:2000yi, Basak:2012bd, Degrassi:2012ry, Zhao:2022ymd, Carena:1995bx}. The mass matrix for neutralinos in the basis $({-i \tilde{B}^0}, {-i \tilde{W}^0}, \tilde{H}_{d}^{0}, \tilde{H}_{u}^{0}, \tilde{S}, \tilde{T}^{0}, \tilde{\bar{T}}^{0})$, where ${\tilde{W}^0}\equiv{\tilde{W}^3}$ is the two-component Weyl notation of the neutral wino, is

\begin{eqnarray}
m_{\tilde{\chi}^{0}} & = & \left(\begin{array}{ccccccc}
M_{1} & 0 & -\frac{1}{2} g_{1} v_{d} & \frac{1}{2} g_{1} v_{u} & 0 & g_{1} v_{T} & -g_{1} v_{\bar{T}} \\
0 & M_{2} & \frac{1}{2} g_{2} v_{d} & -\frac{1}{2} g_{2} v_{u} & 0 & -g_{2} v_{T} & g_{2} v_{\bar{T}} \\
-\frac{1}{2} g_{1} v_{d} & \frac{1}{2} g_{2} v_{d} & \sqrt{2} v_{T} \chi_{d} & -\frac{1}{\sqrt{2}} v_{s} \lambda & -\frac{1}{\sqrt{2}} v_{u} \lambda & \sqrt{2} v_{d} \chi_{d} & 0 \\
\frac{1}{2} g_{1} v_{u} & -\frac{1}{2} g_{2} v_{u} & -\frac{1}{\sqrt{2}} v_{s} \lambda & \sqrt{2} v_{\bar{T}} \chi_{u} & -\frac{1}{\sqrt{2}} v_{d} \lambda & 0 & \sqrt{2} v_{u} \chi_{u} \\
0 & 0 & -\frac{1}{\sqrt{2}} v_{u} \lambda & -\frac{1}{\sqrt{2}} v_{d} \lambda & \sqrt{2} v_{s} \kappa & -\frac{1}{\sqrt{2}} \lambda_{T} v_{\bar{T}} & -\frac{1}{\sqrt{2}} \lambda_{T} v_{T} \\
g_{1} v_{T} & -g_{2} v_{T} & \sqrt{2} v_{d} \chi_{d} & 0 & -\frac{1}{\sqrt{2}} \lambda_{T} v_{\bar{T}} & 0 & -\frac{1}{\sqrt{2}} \lambda_{T} v_{s} \\
-g_{1} v_{\bar{T}} & g_{2} v_{\bar{T}} & 0 & \sqrt{2} v_{u} \chi_{u} & -\frac{1}{\sqrt{2}} \lambda_{T} v_{T} & -\frac{1}{\sqrt{2}} \lambda_{T} v_{s} & 0
\end{array}\right),
\end{eqnarray}
and this matrix is diagonalized by $N$: $N^{*} m_{\tilde{\chi}^{0}} N^{\dag}=m_{\tilde{\chi}^{0}}^{dia}$. The mass matrix for charginos in the basis $\left(-i \tilde{W}^{-},\tilde{H}_{d}^{-},\tilde{T}^{-}\right),\left(-i \tilde{W}^{+},\tilde{H}_{u}^{+},\tilde{T}^{+}\right)$, where $\tilde{W}^{\pm}\equiv\frac{{\tilde{W}^1}\mp i {\tilde{W}^2}}{\sqrt{2}}$ is the two-component Weyl notation of the charged wino, is
\begin{eqnarray}
m_{\tilde{\chi}^-}=\left(\begin{array}{ccc}M_2&\frac{1}{\sqrt{2}}g_2v_u&g_2v_T\\ \frac{1}{\sqrt{2}}g_2v_d&\frac{1}{\sqrt{2}}v_s\lambda&-v_d\chi_d\\ g_2v_{\bar{T}}&-v_u\chi_u&\frac{1}{\sqrt{2}}\lambda_Tv_s\end{array}\right),
\end{eqnarray}
and this matrix is diagonalized by $U$ and $V$: $U^*m_{\tilde{\chi}^{-}}V^{\dagger}=m_{\tilde{\chi}^{-}}^{\rm{diag}}$. There is also a doubly charged triplino $\tilde{\chi}^{++}$ in this model, and its mass is given by $m_{\tilde{\chi}^{++}}=\mu_{T}=\frac{1}{\sqrt{2}} \lambda_{T} v_s$.

\section{Muon anomaly and the LSP relic density} \label{Sec III}

\subsection{Muon anomaly}

\begin{figure}[htb]
\begin{center}
\includegraphics[width=5.25in]{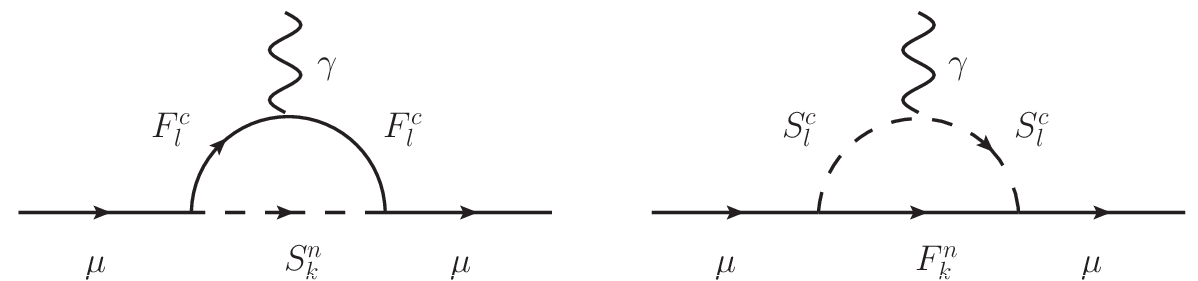}
\caption[]{\label{gm2one} The one-loop Feynman diagrams that contribute the muon $g-2$.  The left diagram represents
the contributions to $a_{\mu}$ from charged fermions $F_l^c$ (e.g. charginos) and neutral scalars $S_k^n$ (e.g. sneutrinos), and the right one represents the contributions from neutral fermions $F_k^n$ (e.g. neutralinos) and charged scalars $S_l^c$ (e.g. sleptons).}
\end{center}
\end{figure}

\begin{figure}[htb]
\begin{center}
\includegraphics[width=6.25in]{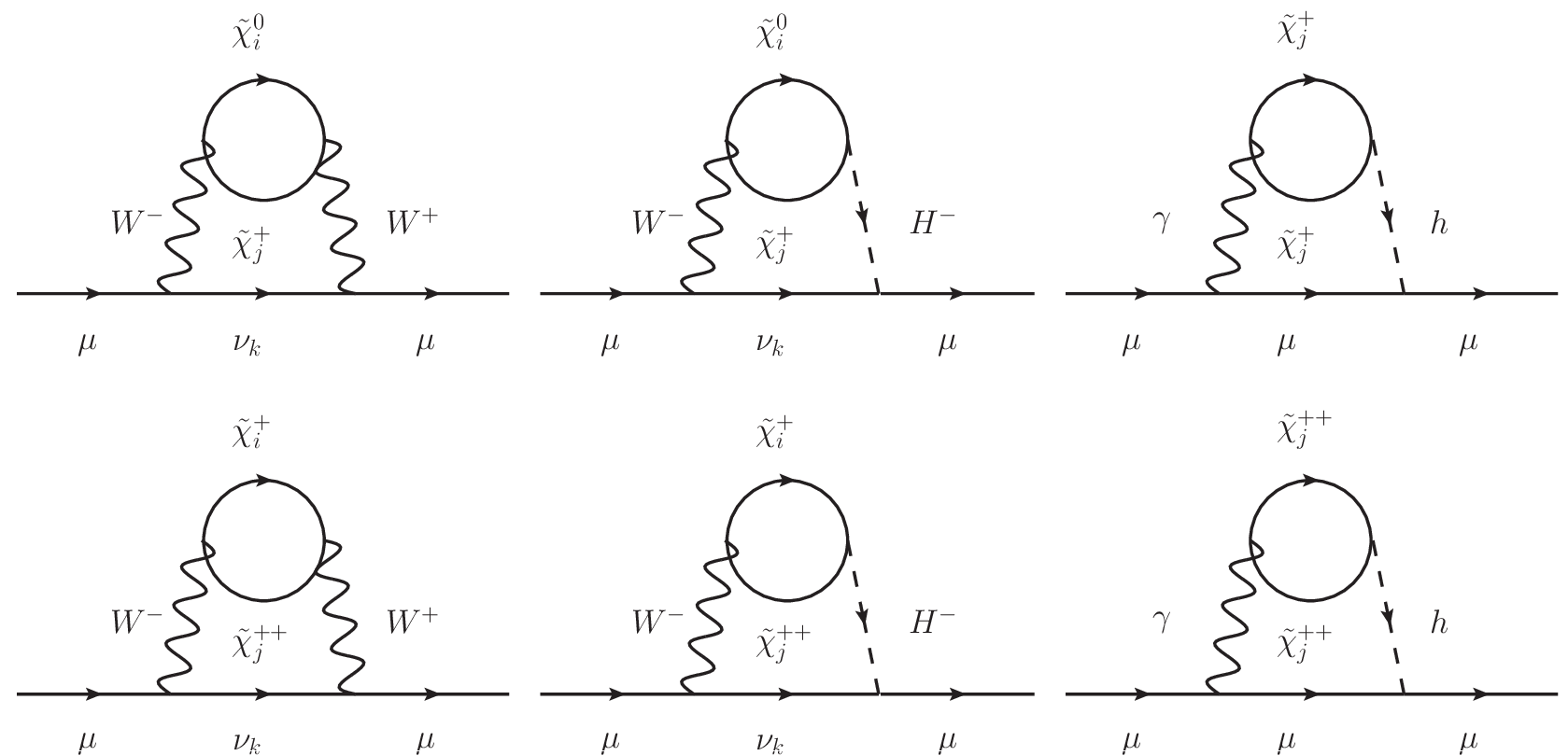}
\caption[]{\label{gm2two} Two-loop Barr-Zee type Feynman diagrams which can contribute the muon $g-2$, the corresponding contributions to $a_{\mu}$ are obtained by attaching a photon to the internal particles in all possible ways.}
\end{center}
\end{figure}

In the TNMSSM, the one-loop NP contributions to the muon anomalous
magnetic moment $a_\mu$ are depicted in Fig.~\ref{gm2one}. The dominant contributions, at one-loop level, come from the chargino-sneutrino loop and the neutralino-charged slepton loop. We present the two-loop Barr-Zee type Feynman diagrams that may have sizable contributions in Fig.~\ref{gm2two}. As can be seen in Fig.~\ref{gm2two}, new Feynman diagrams appear due to the presence of doubly charged fermion $\tilde{\chi}^{++}$. We then can write the NP contribution to $a_\mu$ in the TNMSSM as $\Delta a_\mu^{\mathrm{NP}}=\Delta a_\mu^{\mathrm{one-loop}}+\Delta a_\mu^{\mathrm{two-loop}}$, where the concrete calculations of $\Delta a_{\mu}^{\text{one}-\text{loop}}$ and $\Delta a_{\mu}^{\text{two}-\text{loop}}$ can be found in our previous works~\cite{Feng:2009gn, Zhang:2015csm, Yang:2021duj} (or Ref.~\cite{Moroi:1995yh} for the usual one-loop correction and its result obtained by using the method of Mass Insertion Approximation (MIA)).

\subsection{Standard calculation of relic abundance including coannihilation}

Following Refs.~\cite{Griest:1990kh,Edsjo:2003us}, the calculation of relic abundance including coannihilation is quite different from that of self-annihilation case. In the case of coannihilation involving $N$ species of particles, the masses of these particles are ordered, for convenience, in increasing order, $m_1 \leq m_2 \leq \cdots \leq m_N $, where $m_i$ is the mass of coannihilation particle $\chi_i$ and the corresponding internal degree of freedom, in this convention, is denoted by $g_i$. If the LSP $\chi_1$ is stable and the others can decay into it, all the $\chi_i$ which survive annihilation would eventually decay into $\chi_1$. Therefore, the relic number density $n$ is the sum of the number densities of all $\chi_i$, i.e., $n=\sum\limits_{i=1}^Nn_i$. Using the assumption that $n_i/n\approx n_i^{\rm{eq}}/n^{\rm{eq}}$, where $n^{\rm{eq}}$ is the equilibrium number density, we can write the Boltzmann equation of number density $n$ evolution as
\begin{eqnarray}
\frac{dn}{dt}=-3Hn-\langle\sigma_{\rm{eff}}v\rangle\left(n^2-(n^{\rm{eq}})^2\right),
\end{eqnarray}
where $H$ is the Hubble parameter, and $\langle\sigma_{\rm{eff}}v\rangle$ is the effective thermally-averaged annihilation cross section. In order to express this effective cross section $\sigma_{\rm{eff}}$ more concise, it is convenient to define
\begin{eqnarray}
r_i\equiv n_i^{\rm{eq}}/n^{\rm{eq}}=\frac{g_i(1+\Delta_i)^{3/2}\exp(-x\Delta_i)}{g_{\rm{eff}}},
\end{eqnarray}
where $x=m_1/T$, $\Delta_i=(m_i-m_1)/m_1$ and $g_{\rm{eff}}=\sum\limits_{i=1}^N g_i(1+\Delta_i)^{3/2}\exp(-x\Delta_i)$. The effective cross section then can be expressed as $\sigma_{\rm{eff}} =\sum_{i j}\sigma_{ij}r_ir_j$, where $\sigma_{ij}$ denote the cross section $\sigma (\chi_i \chi_j \rightarrow \text{all possible SM particles})$.

The relic density can be calculated by
\begin{eqnarray}
\Omega h^2 = \frac{1.07\times10^9 \rm{GeV}^{-1}}{g_*(T_{fo})^{1/2} m_{\rm{pl}} J},\label{Omegah2}
\end{eqnarray}
where $m_{\rm{pl}}=1.22\times10^{19}~\rm{GeV}$ is the Plank scale, and $g_*(T_{fo})$ is the total number of effectively relativistic degrees of freedom at the temperature of freeze-out $T_{fo}$, which can be determined by freeze-out condition $\langle\sigma_{\rm{eff}}v\rangle n|_{T=T_{fo}}\approx H|_{T=T_{fo}}$. Defining $x_F\equiv\frac{m_1}{T_{fo}}$, the quantity $J$ in Eq.~(\ref{Omegah2}) then can be expressed by
\begin{eqnarray}
J(x_F)=\int\limits_{x_F}^\infty\frac{\langle\sigma_{\rm{eff}} v\rangle}{x^2}dx.
\end{eqnarray}

\subsection{Direct detection}

The experiment of dark matter direct detection, involving the scattering process $\tilde{\chi}_1^0+q \rightarrow \tilde{\chi}_1^0+q$, can give strong constraint on the SUSY parameter space. Since the squarks in $s$-channel are heavy, the dominant contribution, in this model, to direct detection spin-independent (spin-dependent) scattering cross section $\sigma^{\rm{SI}}$ ($\sigma^{\rm{SD}}$) comes from the exchange of the CP-even Higgs (vector bosons $Z$) in $t$-channel. Based on this analysis, we can write the effective Lagrangian $\mathcal{L}_{\rm{eff}}$ as
\begin{eqnarray}
\mathcal{L}_{\rm{eff}}=C_f^{\rm{SI}}\bar{\tilde{\chi}}_1^0\tilde{\chi}_1^0\bar{f} f+C_f^{\rm{SD}}\overline{\tilde{\chi}^0_1}\gamma_\mu\gamma^5\tilde{\chi}^0_1\bar{f}\gamma^\mu\gamma^5f.
\end{eqnarray}
The WIMP-proton spin-independent scattering cross section $\sigma_{p}^{\rm{SI}}$ then can be expressed as~\cite{Freytsis:2010ne}
\begin{eqnarray}
\sigma_{p}^{\rm{SI}}=\frac{4m_{\tilde{\chi}_1^0}^2m_N^2}{\pi(m_{\tilde{\chi}_1^0}+m_N)^2} \left[Z_{p}f_{p}+(A-Z_{p})f_{n}\right]^{2}/A^2,
\end{eqnarray}
with $Z_p$ denoting the number of proton, $A$ representing the total number of nucleons in an atomic nucleus, and $m_N$ is the mass of the nucleon. The factors $f_{p,n}$ are given by
\begin{eqnarray}
f_{p,n}=\sum\limits_{f=u,d,s}f_{Tf}^{p,n}\frac{C^{\rm{SI}}_{f}m_{N}}{m_{f}}
+\frac{2}{27} \left(1-\sum\limits_{f=u,d,s}f_{Tf}^{p,n}\right) \sum\limits_{f=c,b,t}\frac{C^{\rm{SI}}_{f}m_{N}}{m_{f}},
\end{eqnarray}
where~\cite{Chao:2019lhb}
\begin{eqnarray}
&&f_{Tu}^p=0.0153,f_{Td}^p=0.0191,f_{Ts}^p=0.0447,\nonumber\\
&&f^n_{Tu}=0.0110,f^n_{Td}=0.0273,f^n_{Ts}=0.0447.
\end{eqnarray}
Similarly, the spin-dependent scattering cross section $\sigma_{p,n}^{\rm{SD}}$ reads~\cite{Freytsis:2010ne}
\begin{eqnarray}
\sigma^{\rm{SD}}_{p,n}=\frac{12m_{\tilde{\chi}_1^0}^2m_{N}^2}{\pi(m_{\tilde{\chi}_1^0}
+m_{N})^2} \left[\sum\limits_{f=u,d,s}C^{\rm{SD}}_f\Delta_{f}^{p,n}\right]^2,
\end{eqnarray}
where~\cite{Bringmann:2018lay}
\begin{eqnarray}
\Delta_u^p=\Delta_d^n=0.77,~~\Delta_d^p=\Delta_u^n=-0.47,~~\Delta_s^p=\Delta_s^n=-0.15.
\end{eqnarray}

Note that if the LSP is only a component of total dark matter, the LSP-nucleon scattering cross section should be rescaled by a factor $\Omega_{\rm{LSP}}h^2/0.12$.

\section{Numerical analysis}\label{Sec IV}
In this Section, we discuss the numerical results of $a_\mu$ as well as the relic density and direct detection of the dark matter in TNMSSM. The SUSY particle spectrum is generated by using the generator $\verb"SPheno-4.0.5"$~\cite{Porod:2011nf,Porod:2003um}, and the parameters are given at the low SUSY scale with the renormalization scale $M=1~\rm{TeV}$. This code is numerically reliable and, therefore, widely used in study of SUSY phenomenology.

The SM input parameters we used are $m_W=80.377~\rm{GeV}$, $m_Z=91.1876~\rm{GeV}$, $\sin^2\theta_W(m_Z)=0.2312$, $\alpha_{\rm{em}}(m_Z)=1/128.9$, $\alpha_{\rm{s}}(m_Z)=0.118$ and the fermion coupling constant $G_F=1.1664\times10^{-5}~\rm{GeV}^{-2}$. The masses of SM fermions are taken as $m_e=0.511~\rm{MeV}$, $m_\mu=0.105~\rm{GeV}$, $m_\tau=1.78~\rm{GeV}$, $m_u=2.3~\rm{MeV}$, $m_c=1.28~\rm{GeV}$, $m_t=173.5~\rm{GeV}$, $m_d=4.80~\rm{MeV}$, $m_s=0.095~\rm{GeV}$ and $m_b=4.65~\rm{GeV}$ from the PDG~\cite{ParticleDataGroup:2022pth}.

For the SUSY input parameters, we take the trilinear Higgs slepton couplings $T_e\equiv$$\rm{diag}$ $(y_{e_1},y_{e_2},y_{e_3})A_e$ with $A_e=0~\rm{TeV}$ (in this case the mixing is significant only for the scalar taus), and the trilinear Higgs squark couplings are similarly defined as $T_u\equiv$$\rm{diag}$$(y_{u_1},y_{u_2},y_{u_3})A_q$, $T_d\equiv$$\rm{diag}$$(y_{d_1},y_{d_2},y_{d_3})A_q$ with $A_q=0.1~\rm{TeV}$, where all the $y_i$ are the corresponding Yukawa coupling constants in the superpotential. The masses of squarks are set to be as large as $2~\rm{TeV}$ due to that no direct observation of supersymmetric particles has been made in experiments such as the LHC. Similarly, the masses of $\tilde{e}_{1,2}$ and $\tilde{\tau}_{1,2}$ (the mass eigenstates of scalar electrons and scalar taus) are set to be as large as $2~\rm{TeV}$, and that of gluinos are set to be $3~\rm{TeV}$. The SUSY soft breaking parameters $T_\lambda=0.5~\rm{TeV}$, $T_{\lambda_T}=0.1~\rm{TeV}$, $T_\kappa=-0.1~\rm{TeV}$, $T_{\chi_u}=-0.1~\rm{TeV}$ and $T_{\chi_d}=0.1~\rm{TeV}$ are reasonably fixed, and the rest of parameters would be discussed in the following analysis.

Considering the tree-level masses of $W$ boson and $Z$ boson, in the TNMSSM, the parameter $\rho$ can be written as
\begin{eqnarray}
\rho =\frac{m_{W}^2}{\cos^2\theta_W m_{Z}^2}=1-\frac{2v_{T\bar{T}}^2}{v_{ud}^2},
\end{eqnarray}
where $\theta_{W}$ is the Weinberg angle. Experimentally, the constraint on this parameter within $2\sigma$-deviation requires $\rho_{\rm{exp}}=1.00038\pm0.00020\times2$~\cite{ParticleDataGroup:2022pth}. Therefore, the scenario in which the value of $v_{T\bar{T}}$ does not exceed $1~\rm{GeV}$ is safer. Henceforth, we deem this condition a necessary requirement for a viable parameter space.

Given that the measured SM-like Higgs boson mass is $m_h^{\rm{exp}}=125.25\pm0.17~\rm{GeV}$~\cite{ParticleDataGroup:2022pth}, this would give an important constraint on the parameter space of NP models. As mentioned above, we consider its theoretical value $m_h$ up to two-loop level, and then require its magnitude to be its measured mass within $3 \sigma$ deviation. Using the mass spectrum generator $\verb"SPheno-4.0.5"$, we scan the following parameter space:
\begin{eqnarray}
&&\lambda,\kappa,\lambda_T,\chi_u,\chi_d\in[0.01,1],\nonumber\\
&&\tan\beta\in[2,50],v_{T},v_{\bar{T}}\in[0.01,1]~\rm{GeV},\nonumber\\
&&v_s\in[10^3,10^6]~\rm{GeV},M_1,M_2\in[100,5000]~\rm{GeV},\nonumber\\
&&m_{\tilde{l},{22}}=m_{\tilde{e},{22}}\in[100,2000]~\rm{GeV}.\label{scanvevs}
\end{eqnarray}
Here we have assumed that the soft breaking parameters of smuon $m_{\tilde{l},{22}}$ and $m_{\tilde{e},{22}}$ are identical. For simplicity, we use a new dimension-1 symbol $m_{\tilde{\mu}_{LR}}$($\equiv m_{\tilde{l},{22}}=m_{\tilde{e},{22}}$) rather than either of them to analyse the numerical results in the following discussion. After the scanning of parameter space, it turns out that the most samples that satisfy this $3 \sigma$ constraint are in the region: $0<\lambda<\kappa<1$ and $0<\chi_u<0.5$. Under the large $\tan\beta$ assumption, the tree-level mass is approximately behaved as $m_{h}^2|_{\rm{tree}}\sim m_Z^2-\frac{\lambda^4}{\kappa^2} \frac{v_{u}^2}{2}+\chi_u^2 \frac{v_{u}^2}{2}$~\cite{Agashe:2011ia}, so that the viable region can be understood without difficulty.

Additionally, the rare decay processes of $B$ meson, such as $\bar{B}\to X_s\gamma$ and $B_s^0\to\mu^+\mu^-$, are important for researching the NP models beyond the SM, since their theoretical predictions are strictly constrained by the precise measurements. The average experimental bound on the branching ratios of $\bar{B}\to X_s\gamma$ and $B_s^0\to\mu^+\mu^-$ are~\cite{ParticleDataGroup:2022pth}
\begin{eqnarray}
Br(\bar{B}\to X_s\gamma)&=&(3.49\pm0.19)\times10^{-4},\nonumber\\
Br(B_s^0\to\mu^+\mu^-)&=&(3.01\pm0.35)\times10^{-9}.
\end{eqnarray}
We take both of them into account in our numerical analysis. For more details about these rare decays, please see our previous works~\cite{Yang:2018fvw, Chen:2023zfl} and references therein.

Thus far, we have clarified the scanned range for all parameters as shown in Eq.~(\ref{scanvevs}). Next, we will investigate the scenario where bino LSP serves as the candidate for cold dark matter. As in the MSSM, there would be too large dark matter relic density in the absence of coannihilation mechanism and, therefore, we consider the coannihilation processes that can be effective in this model. Different from the MSSM, there are triplinos that can coannihilate with the bino LSP, so that we mainly focus on this triplino NLSPs coannihilation scenario. Since the effective $\mu$ and $\mu_T$ are closely related to the masses of neutralino/chargino and the doubly charged triplino, it is necessary to discuss the constraints imposed on them.
Based on the analysis above, we only consider the scanned samples which can satisfy the following conditions:
\begin{eqnarray}
&&v_{T\bar{T}}\in[10^{-2},1]~\rm{GeV},\nonumber\\
&&M_1\in[100,500]~\rm{GeV},\mu_T\in[1.1M_1,1.5M_1],\nonumber\\
&&M_2\in[1.5M_1,5000~\rm{GeV}],\mu\in[1.5M_1,5000~\rm{GeV}],\nonumber\\
&&m_{\tilde{\mu}_{LR}}\in[1.5M_1,2000~\rm{GeV}],\nonumber\\
&&m_{scalar}> 1500~\rm{GeV},\label{scan}
\end{eqnarray}
here $m_{scalar}$ represents the mass of the lightest SUSY scalar Higgs (it may be neutral or charged).
Given that the thermal equilibrium number density $n^{\rm{eq}}$, in the non-relativistic approximation, is expressed as $n^{\rm{eq}}\approx g(mT/2\pi)^{3/2}\exp(-m/T)$, the minimum values of $M_2$ and $\mu$ we considered in Eq.~(\ref{scan}) are $1.5 M_1$ and, therefore, only the triplinos coannihilation processes are needed to be taken into account. We also assume that $m_{scalar}> 1500~\rm{GeV}$ and, consequently, the possibility of $scalar$-$pole$ annihilation, $m_{scalar}\sim2m_{\tilde{\chi}^0_1}$, are explicitly not included.

\begin{figure}[htb]
\begin{center}
\begin{minipage}[c]{0.48\textwidth}
\includegraphics[width=3in]{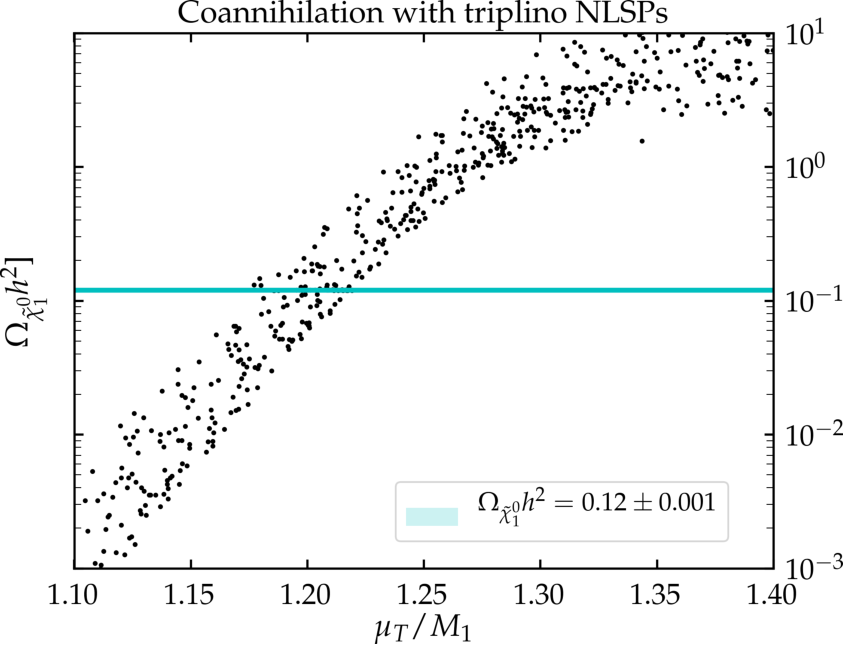}
\end{minipage}%
\begin{minipage}[c]{0.48\textwidth}
\includegraphics[width=3in]{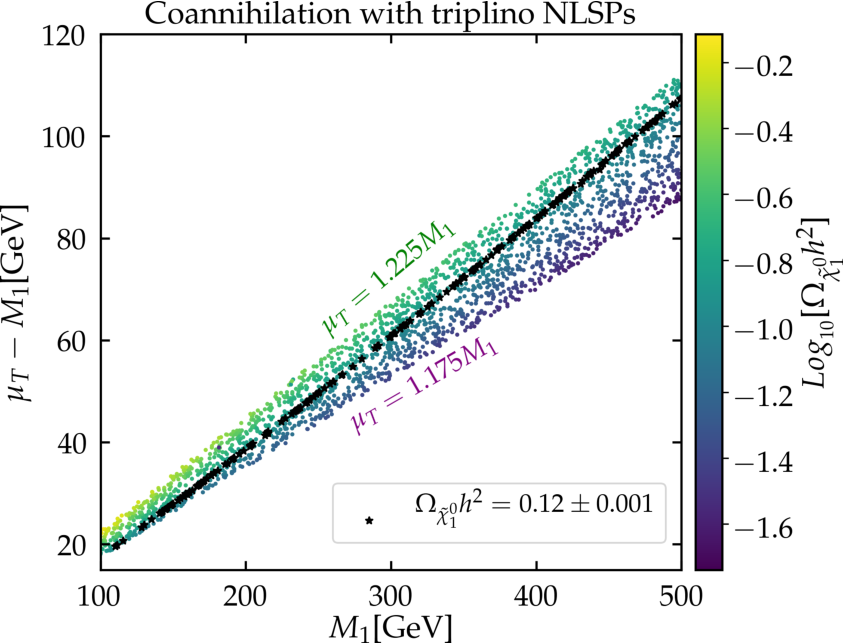}
\end{minipage}
\caption[]{\label{coannihilation} Some sampling points on the planes of dark matter relic density $\Omega_{\tilde{\chi}^0_1}h^2$ versus the ratio of triplino mass to that of bino $\mu_T/M_1$ (left panel) and $\mu_T-M_1$ versus $M_1$ (right panel). The light blue region in left panel represents the observed dark matter relic density within $1\sigma$. In the right panel, all points are colored with various hues according to the value of $\Omega_{\tilde{\chi}^0_1}h^2$. In order to make a distinction between those points satisfying $\Omega_{\tilde{\chi}^0_1}h^2=0.12\pm0.001$ and the others, we highlight them by black star.}
\end{center}
\end{figure}

Firstly, we investigate the coannihilation effect under the assumption that all the triplinos are the NLSPs. With the triplino mass $\mu_T$ is in the region $[1.1M_1,1.5M_1]$, the dominant self-annihilation and coannihilation processes which we need take into account in calculation are
\begin{eqnarray}
&&\text{(1)} :\tilde{\chi}_{j}^{0}\tilde{\chi}_{k}^{0}\to\bar{u}_{i}u_{i},\bar{d}_{i}d_{i},\bar{l}_{i}l_{i},\bar{\nu}_{i}\nu_{i},WW,ZZ,hh,hZ(i=1,2,3,j,k=1,2,3),\nonumber\\
&&\text{(2)} :
\tilde{\chi}_{j}^{0}\tilde{\chi}_{k}^{\pm}\to\bar{u}_{i}d_{i},\bar{l}_{i}\nu_{i},ZW,hW,\gamma W(i=1,2,3,j=1,2,3,k=1),\nonumber\\
&&\text{(3)} :
\tilde{\chi}_{j}^{0}\tilde{\chi}_{k}^{\pm\pm}\to WW(j=1,2,3,k=1),\nonumber\\
&&\text{(4)} :\tilde{\chi}_{j}^{\pm}\tilde{\chi}_{k}^{\pm}\to\bar{u}_{i}u_{i},\bar{d}_{i}d_{i},\bar{l}_{i}l_{i},\bar{\nu}_{i}\nu_{i},WW,ZZ,hh,\gamma\gamma,hZ,\gamma Z,\gamma h(i=1,2,3,j,k=1),\nonumber\\
&&\text{(5)} :\tilde{\chi}_{j}^{\pm}\tilde{\chi}_{k}^{\pm\pm}\to\bar{u}_{i}d_{i},\bar{l}_{i}\nu_{i},ZW,hW,\gamma W(i=1,2,3,j,k=1),\nonumber\\
&&\text{(6)} :\tilde{\chi}_{j}^{\pm\pm}\tilde{\chi}_{k}^{\pm\pm}\to\bar{u}_{i}u_{i},\bar{d}_{i}d_{i},\bar{l}_{i}l_{i},\bar{\nu}_{i}\nu_{i},WW,ZZ,hh,\gamma\gamma,hZ,\gamma Z,\gamma h(i=1,2,3,j,k=1),\nonumber
\end{eqnarray}
and the corresponding Feynman diagrams are listed in Fig.~\ref{co1-1}-\ref{co5-2} of the Appendix~\ref{APPENDIX B}. The scan results are plotted in Fig.~\ref{coannihilation}, all the sampling points are survivable under the constraints from SM-like Higgs mass within $3\sigma$ and $\bar{B}\to X_s\gamma$ or $B_s^0\to\mu^+\mu^-$ within $2\sigma$.

In Fig.~\ref{coannihilation}, the left panel shows that the theoretical relic density of dark matter versus the ratio of triplino mass to that of bino $\mu_T/M_1$. As can be seen clearly, at a small value of $\mu_T/M_1$, such as $1.15$, the relic density is too small to explain the Planck observation by itself, and other dark matter candidates are needed. As the ratio increases, the coannihilation effect becomes weak due to the exponentially suppressed number density of NLSPs, and an excessively large relic density would reemerge with a large enough ratio such as $1.35$. It then turns out that the ratio $\mu_T/M_1$ needs to stay in the interval $[1.175,1.225]$ roughly for bino dark matter being responsible for the total dark matter relic density $0.12\pm0.001$, and we also show this interval in more detail in the right panel of Fig.~\ref{coannihilation}. From the right panel, we can find that all of the bino dark matter with a mass in the region $[100,500]~\rm{GeV}$ can give a relic density satisfying the Planck constraint $\Omega_{\rm{DM}}h^2=0.12\pm0.001$. Additionally, the ratio needs to be mildly changed for different bino mass $M_1$, and a larger ratio is required for a heavier bino-like dark matter candidate.

Secondly, we study the constraints from $a_\mu$, the LHC and direct detection of WIMP on the parameter space together.

Given that the deviation between the experiment and the SM prediction of $a_\mu$ has increased (from $4.2\sigma$ to $5.1\sigma$ after combining the Run-2 and Run-3 data of the FNAL Muon $g-2$ experiment), this bound is so significant that it can not be treated loosely. In our analysis, we probe the parameter space that can accommodate the NP contribution $\Delta a_\mu^{\rm{NP}}$ within $2\sigma$.

Since the electroweakinos and sleptons can potentially be light with sufficiently large mass gap ($\geq1.5 m_{\tilde{\chi}^0_1}$) with the bino-like LSP, there are also strict limits on the masses of winos and sleptons from the LHC search in pair production of electroweakinos and sleptons. Here we concisely summarize the LHC results\footnote{In this work, we adopt a somewhat crude yet effective approach by directly applying these LHC limits to our parameter space to discuss their implications. It is also worth noting that LHC limits are usually interpreted in terms of ``simplified models"\cite{LHCNewPhysicsWorkingGroup:2011mji} with specific assumptions on the mass and coupling configurations, and the sensitivity of the searches may change significantly if there are deviations from these assumptions\cite{Chakraborti:2020vjp}. In cases where high precision is crucial, before applying these limits to the parameter space of theoretical models, they should be reinterpreted with tools such as CheckMATE\cite{Drees:2013wra,Kim:2015wza,Dercks:2016npn}, SModelS\cite{Kraml:2013mwa,Kraml:2014sna}, etc. This is indeed the standard procedure for applying these limits. Nevertheless, achieving such high precision is not what we are aiming for, and a rough approximation is adequate for our analysis.} that are most relevant for our analysis as follows:
\begin{itemize}
    \item \textbf{Limit-1} \cite{ATLAS:2019lff} and  \textbf{Limit-2} \cite{ATLAS:2018ojr}: The limit from the searches of slepton pair production in the dilepton final state $\tilde{l}^{+} \tilde{l}^{-} \rightarrow\left(l^{+} \tilde{\chi}_{1}^{0}\right)\left(l^{-} \tilde{\chi}_{1}^{0}\right) \rightarrow 2 l+{E}_T^{\rm{miss}}$, where ${E}_T^{\rm{miss}}$ represents the missing energy.
    \item \textbf{Limit-3} \cite{ATLAS:2019lng}: The limit from the searches of slepton pair production in the dilepton final state with the presence of initial-state radiation (ISR) $\tilde{l}^+ \tilde{l}^- \rightarrow (l^+ \tilde{\chi}_1^0)(l^- \tilde{\chi}_1^0) \rightarrow 2l + {E}_T^{\rm{miss}} + \text{ISR}$.
    \item \textbf{Limit-4} \cite{ATLAS:2020pgy}: The bound from searches performed by ATLAS at 139 $\rm{fb}^{-1}$ which look for decay of electroweakinos pairs through on-shell $W$ and Higgs bosons $\tilde{\chi}_{\rm{wino-like}}^{\pm} \tilde{\chi}_{\rm{wino-like}}^0 \rightarrow (W \tilde{\chi}_1^0)(h \tilde{\chi}_1^0) \rightarrow l + b \bar{b} + {E}_T^{\rm{miss}}$.
    \item \textbf{Limit-5} \cite{ATLAS:2018ojr}: The bound from the ATLAS searches which look for decays of electroweakinos pairs through on-shell gauge bosons in the 3 lepton final state $\tilde{\chi}_{\rm{wino-like}}^{\pm} \tilde{\chi}_{\rm{wino-like}}^0 \rightarrow (W \tilde{\chi}_1^0)(Z \tilde{\chi}_1^0) \rightarrow 3l + {E}_T^{\rm{miss}}$ and in the 2 lepton final state with jets $\tilde{\chi}_{\rm{wino-like}}^{\pm} \tilde{\chi}_{\rm{wino-like}}^0 \rightarrow (W \tilde{\chi}_1^0)(Z \tilde{\chi}_1^0) \rightarrow 2l + \text{jets} + {E}_T^{\rm{miss}}$. There is also a bound from the decay of two charginos $\tilde{\chi}_{\rm{wino-like}}^{+} \tilde{\chi}_{\rm{wino-like}}^{-} \rightarrow (W^+ \tilde{\chi}_1^0)(W^- \tilde{\chi}_1^0) \rightarrow 2l + {E}_T^{\rm{miss}}$. However, it is much weaker than the first two, so we do not show it in the following analysis.
    \item \textbf{Limit-6} \cite{ATLAS:2019lng}: The bound from searches which look for decay of electroweakinos pairs through off-shell gauge bosons in the 2 lepton final state with the ISR $\tilde{\chi}_{\rm{wino-like}}^{\pm} \tilde{\chi}_{\rm{wino-like}}^0 \rightarrow (W^{*} \tilde{\chi}_1^0)(Z^{*} \tilde{\chi}_1^0) \rightarrow 2l + {E}_T^{\rm{miss}} + \text{ISR}$.
    \item \textbf{Limit-7} \cite{ATLAS:2018ojr}: The constraint from pair production of electroweakinos decaying via the processes $\tilde{\chi}_{\rm{wino-like}}^{\pm} \tilde{\chi}_{\rm{wino-like}}^0 \rightarrow (\tilde{l}^{\pm} \nu)(\tilde{l}^{+} l^{-}) \rightarrow 3l + {E}_T^{\rm{miss}}$ and $\tilde{\chi}_{\rm{wino-like}}^{\pm} \tilde{\chi}_{\rm{wino-like}}^0 \rightarrow ({l}^{\pm} \tilde{\nu})(\tilde{l}^{+} l^{-}) \rightarrow 3l + {E}_T^{\rm{miss}}$ to $3l$ final state. There are also limit from the decay processes of two charginos $\tilde{\chi}_{\rm{wino-like}}^{+} \tilde{\chi}_{\rm{wino-like}}^{-} \rightarrow (\tilde{l}^{+} \nu)(\tilde{l}^{-} {\nu}) \rightarrow 2l + {E}_T^{\rm{miss}}$ and $\tilde{\chi}_{\rm{wino-like}}^{+} \tilde{\chi}_{\rm{wino-like}}^{-} \rightarrow ({l}^{+} \tilde{\nu})({l}^{-} \tilde{\nu}) \rightarrow 2l + {E}_T^{\rm{miss}}$ to $2l$ final state. However, the $3l$ final state constraint is much strict and we only show it in the subsequent analysis.
\end{itemize}

For the constraints from direct detection of WIMP, the most stringent bound comes from the spin-independent (SI) WIMP-nucleon scattering, and the latest bound on this scattering cross section is given by the LUX-ZEPLIN (LZ) experiment \cite{LZ:2022lsv}. Except for this bound, there would be more strict constraints in the near future, such as the Xenon-nT projection (For a 20 t y exposure, a $50~\rm{GeV}/c^2$ WIMP with cross-section of $2.6 \times 10^{-48} \rm{cm}^2 (5.0 \times 10^{-48} \rm{cm}^2)$ will yield a median $3\sigma$ ($5\sigma$) discovery significance \cite{XENON:2020kmp}). Therefore, we take these direct detection constraints into account in the following analysis.

\begin{figure}[htb]
\begin{center}
\begin{minipage}[c]{0.48\textwidth}
\includegraphics[width=3.1in]{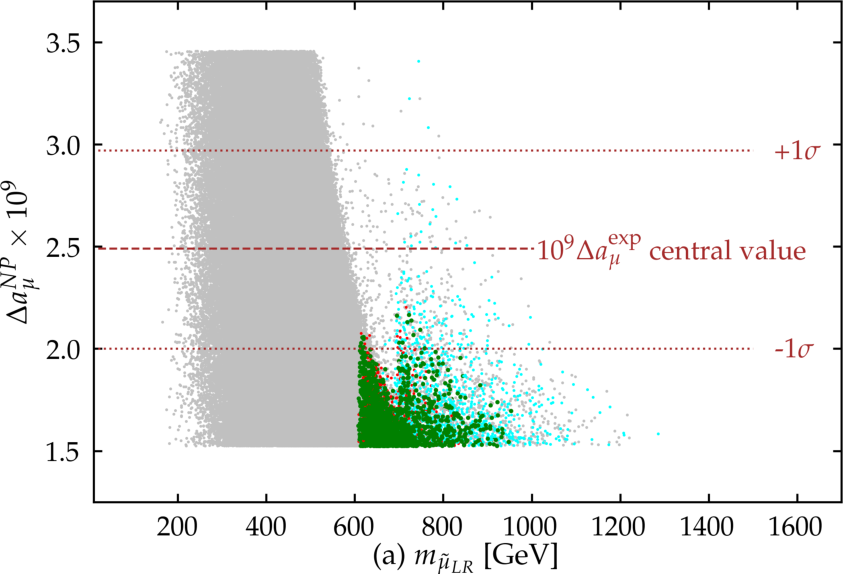}
\end{minipage}
\begin{minipage}[c]{0.48\textwidth}
\includegraphics[width=3.1in]{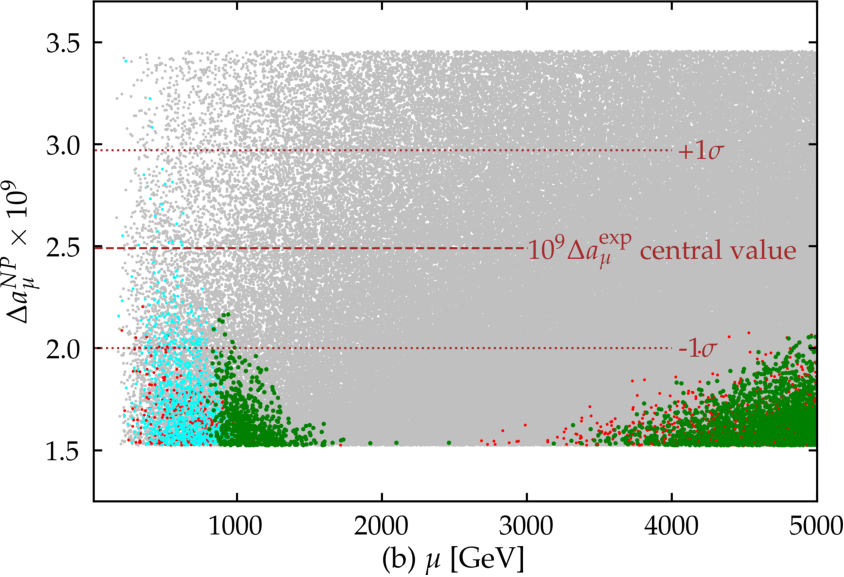}
\end{minipage}\\
\begin{minipage}[c]{0.48\textwidth}
\includegraphics[width=3.1in]{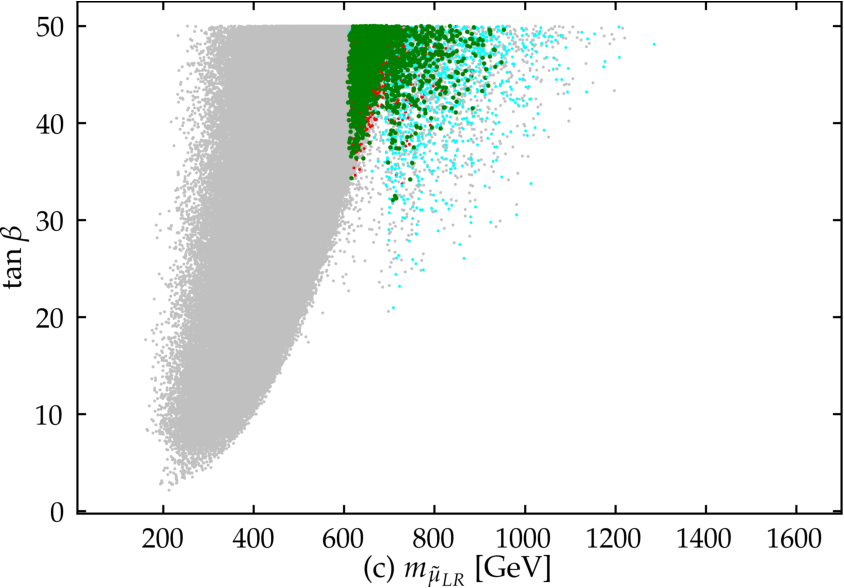}
\end{minipage}
\begin{minipage}[c]{0.48\textwidth}
\includegraphics[width=3.1in]{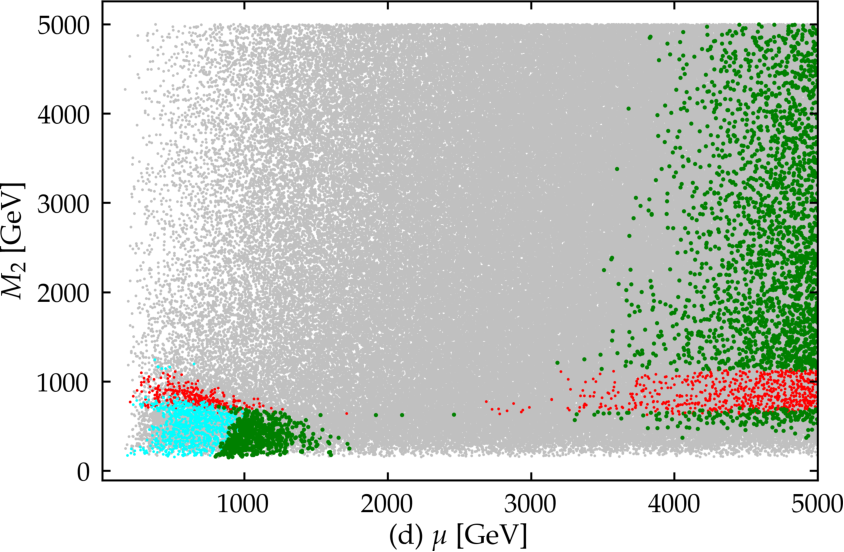}
\end{minipage}%
\caption[]{\label{fig:gm21234} The constraints on the parameter space from the latest $a_\mu$ measurements, LHC and LZ-2022 direct detection experiment \cite{LZ:2022lsv}. The central value and $\pm 1\sigma$ of $\Delta a_\mu^{\rm{exp}}$ are shown in (a) and (b). For the color coding: please see text.}
\end{center}
\end{figure}

\begin{figure}[!htb]
\begin{center}
\begin{minipage}[c]{0.48\textwidth}
\includegraphics[width=3.1in]{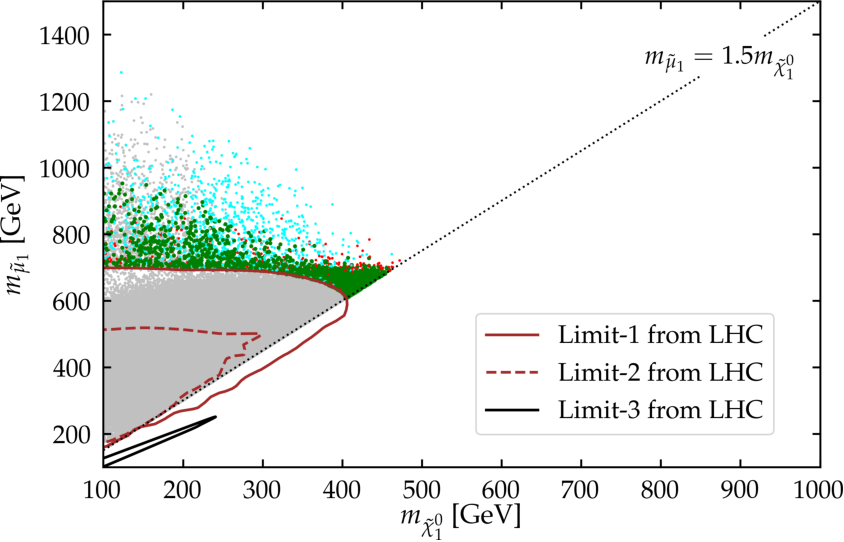}
\end{minipage}%
\begin{minipage}[c]{0.48\textwidth}
\includegraphics[width=3.1in]{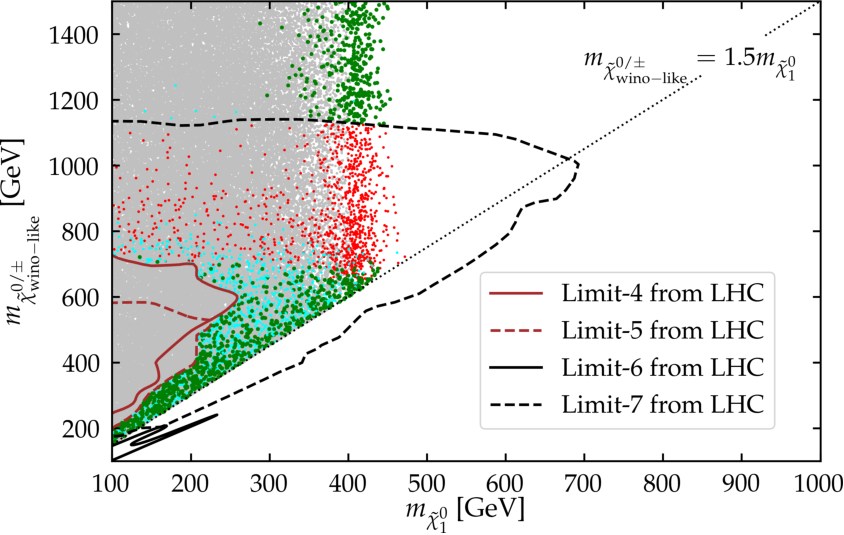}
\end{minipage}
\caption[]{\label{LHC} Same as the samples of Fig.~\ref{fig:gm21234}, but showing $m_{\tilde{\mu}_{1}}$ (left pannel) and $m_{\tilde{\chi}^{0/\pm}_{\rm{wino-like}}}$ (right pannel) versus the mass of LSP $m_{\tilde{\chi}^0_1}$ respectively. The LHC bounds mentioned in text are shown. For the Limit-7, we only take it into account if $m_{\tilde{\mu}_{1}}$ is lighter than $m_{\tilde{\chi}^{0/\pm}_{\rm{wino-like}}}$ due to that the slepton involved in the corresponding decay processes is required to be on-shell. For the color coding: see text.}
\end{center}
\end{figure}

\begin{figure}[htb]
\begin{center}
\begin{minipage}[c]{0.48\textwidth}
\includegraphics[width=3.1in]{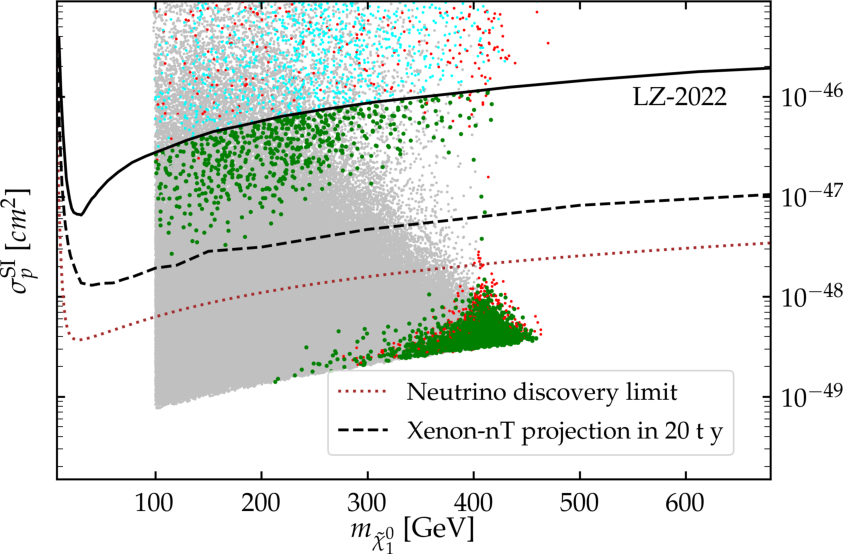}
\end{minipage}
\caption[]{\label{DD SI}Same as the samples of Fig.~\ref{fig:gm21234}, this figure shows the WIMP-proton spin-independent scattering cross section $\sigma_{p}^{\rm{SI}}$ versus mass of the bino-like LSP $m_{\tilde{\chi}^0_1}$. The LZ-2022 bound is plotted by the black solid line. The future experimental sensitivity (here we take the Xenon-nT projection in 20 t y \cite{XENON:2020kmp}) and the neutrino discovery limit \cite{Billard:2013qya} are also shown in this figure by black dashed and brown dotted lines, respectively. For the color coding: see text.}
\end{center}
\end{figure}

In respect to all the constraints mentioned above, we plot the scan results in Fig.~\ref{fig:gm21234} ($a_\mu$ bound), Fig.~\ref{LHC} (LHC limits) and Fig.~\ref{DD SI} (direct detection constraints). All the sampling points we plotted can reproduce the SM-like Higgs mass within $3\sigma$, and they also satisfy the constraints from $\bar{B}\to X_s\gamma$ or $B_s^0\to\mu^+\mu^-$ within $2\sigma$, as well as $\Omega_{\tilde{\chi}_1^0} h^2$ within $1\sigma$. The points denoted with different colors in these figures survive certain constraints:
\begin{itemize}
    \item \textbf{gray}: points that survive $a_\mu$ constraint within $2\sigma$ but are excluded by Limit-1 - Limit-6 from LHC.
    \item \textbf{red}: points that are in agreement with $a_\mu$ within $2\sigma$ and not excluded by Limit-1 - Limit-6 but excluded by Limit-7 from LHC. For the Limit-7, we only take it into account if $m_{\tilde{\mu}_{1}}$ is lighter than $m_{\tilde{\chi}^{0/\pm}_{\rm{wino-like}}}$ due to that the slepton involved in the corresponding decay processes is required to be on-shell.
    \item \textbf{cyan}: points that additionally pass the Limit-7 from LHC but can not fulfill the direct detection constraint.
    \item \textbf{green}: points that can pass all the constraints considered above.
\end{itemize}

Before further discussion, we now comment on the NP contribution to $a_\mu$. At the one-loop level, one might expect the contribution to $a_{\mu}$ in TNMSSM to be quite different from that in MSSM since there are many newly added particles compared to the MSSM (Higgses of the triplets/singlet and their supersymmetric partners). Unfortunately, this is not the case because these newly added gauge eigenstates, in $a_{\mu}$ calculation via the method of MIA, do not appear at all in the one-loop Feynman diagrams due to the fact that the neutrino mass term $\xi \hat{l} \cdot \hat{T} \hat{l}$ is not introduced in the considered version of TNMSSM. In addition, even though the term $\xi \hat{l} \cdot \hat{T} \hat{l}$ is added to the potential to obtain the tiny neutrino masses as mentioned in the $Introduction$ section, the small dimensionless parameter $\xi$ (of the order of $m_{\nu_\mu}/v_T$, where $m_{\nu_\mu}$ is the mass of neutrino $\nu_\mu$) would lead to the contribution to $a_{\mu}$ from these newly added particles being negligibly small (roughly estimated to be less than $0.001\%\Delta a_\mu^{\rm{one-loop}}$). Consequently, the one-loop contribution to $a_\mu$ from the new particles can be safely omitted. However, these new particles can make contributions to $\Delta a_\mu^{\mathrm{NP}}$ through the considered Barr-Zee type two-loop diagrams as shown in Fig. \ref{gm2two}, and these contributions are comparable with the ones from the MSSM particles. For the two-loop contribution, the Barr-Zee type of Feynman diagrams would contribute a $\Delta a_\mu^{\mathrm{two-loop}}\sim10^{-10}$, which corresponds to about $3\%-5\%~\Delta a_\mu^{\mathrm{NP}}$. Therefore, the NP contribution to $a_\mu$ is mainly from the one-loop MSSM contribution, and we can just treat $a_\mu$ constraint as a requirement that viable parameter space must fulfill.


From Fig.~\ref{fig:gm21234} (a) - (d), it is evident that there are two distinguishable regions consistent with the $a_\mu$ constraint within $2\sigma$. To clarify this pattern, we now temporarily focus on the colored points (red, cyan and green) only. Within the region where $\mu$ is less than $2~\rm{TeV}$, these sampled points exhibit approximate symmetry around the line $M_2 = \mu$ (see Fig.~\ref{fig:gm21234} (d)). This can be easily understood if one calculates $a_\mu$ by using the method of MIA. In this scenario, the dominant contribution to $a_\mu$ arises from the loop involving $\tilde{W}^{\pm}-\tilde{H}^{\pm}-\tilde{\nu}_\mu$ and, according to MIA, the analytical result of this contribution remains unchanged when the values of $M_2$ and $\mu$ are exchanged \cite{Athron:2015rva}. Therefore, the shape appears without unnaturalness. Similarly, in the region where $\mu$ is larger than $2~\rm{TeV}$, the dominated contribution to $a_\mu$ can also be figured out, without difficulty, to be bino contribution from ${\tilde{B}^0}-\tilde{\mu}_{L}-\tilde{\mu}_R$ loop. The most obvious feature for this bino-dominated scenario is that the NP contribution is approximately proportional to the parameter $\mu$ \cite{Athron:2015rva}, and this is exactly what Fig.~\ref{fig:gm21234} (b) shows. Based on these analysis, the majority of gray points clearly belong to the bino-dominated scenario (see Fig.~\ref{fig:gm21234} (a) or (c)), and they are excluded by the Limit-1 - Limit-3 from the LHC search in pair production of sleptons, which requires the lightst smuon to have a mass heavier than about $600~\rm{GeV}$ (see Fig.~\ref{LHC} left panel). All the rest of gray points, where the dominated contribution to $a_\mu$ comes from $\tilde{W}^{\pm}-\tilde{H}^{\pm}-\tilde{\nu}_\mu$ loop, are excluded mainly by the Limit-4 - Limit-6 from the LHC search in pair production of electroweakinos (see Fig.~\ref{LHC} right panel). The red points are ruled out by the LHC Limit-7 as shown in Fig.~\ref{LHC} right panel, and this limit is active only for that $m_{\tilde{\mu}_{1}}$ is smaller than $m_{\tilde{\chi}^{0/\pm}_{\rm{wino-like}}}$. Roughly estimated, most samples with $M_2$ in the range [700,1150] $\rm{GeV}$ are excluded by this constraint. The cyan points, which are excluded by the LZ-2022 constraint shown in Fig.~\ref{DD SI}, indicate that the higgsino mass $\mu$ needs to be at least about $750~\rm{GeV}$ (see Fig.~\ref{fig:gm21234} (b)). As the scalar quarks in the s-channel are heavy (as large as 2.0 $\rm{TeV}$), the dominant contribution to $\sigma^{\rm{SI}}$ comes from the exchange of the SM-like Higgs in t-channel. The coupling of $\tilde{\chi}^0_1-\tilde{\chi}^0_1-h$ originates from gaugino Yukawa coupling of the form $h^\dagger \tilde{H}^0 {\tilde{B}^0}$, so that a small mix between $\tilde{H}^0$ and ${\tilde{B}^0}$ is required. This coincides with that points with small $\mu$ are ruled out. Additionally, the parameter space corresponding to the bino-dominated scenario is not constrained by LZ-2022 due to the sufficiently large $\mu$ parameter.

Finally, the green points, plotted in Fig.~\ref{fig:gm21234} - \ref{DD SI}, satisfy all the aforementioned constraints. A $\tan \beta$ larger than, at least, 30 is required if one expects a $a_\mu$ within $2\sigma$. The NP contribution to $a_\mu$ can satisfy the constraint within $1\sigma$ if a large enough $\tan \beta$ (as large as about 50) is allowed. The permissible mass range for bino-like LSP $m_{\tilde{\chi}^0_1}$ is between 100 and 450 $\rm{GeV}$. For the green points in the bino-dominated scenario, the allowed $m_{\tilde{\chi}^0_1}$ is in the range [200,450] $\rm{GeV}$. There is almost no prospect of direct detection in the future, as these points fall below the neutrino discovery limit. However, this region may be tested if future LHC searches impose stricter limits on the mass of the smuon in pair production of sleptons. The rest of green points indicate that the LSP mass $m_{\tilde{\chi}^0_1}$ should be in the range [100,400] $\rm{GeV}$, with lighter LSP more easily escaping current experimental limitation. Additionally, all of them can be probed in the near future experiments such as Xenon-nT projection in 20 t y, as explicitly shown in Fig.~\ref{DD SI}.
\section{Summary}\label{Sec V}
In this paper, we focus on the scenario that bino-like LSP is the dark matter candidate which can give a correct relic density via coannihilation with the triplino NLSPs in the TNMSSM. The SM-like Higgs mass at two-loop level within $3\sigma$ has been reproduced, and the limitations from rare decay processes of $B$ meson $\bar{B}\to X_s\gamma$ and $B_s^0\to\mu^+\mu^-$ are satisfied in the assumed parameter space. The muon anomalous magnetic moment at two-loop level, dark matter relic density involving coannihilation and the direct detection spin-independent scattering cross sections are carefully calculated. Considering the latest limitations from the muon anomaly $a_\mu$, LHC and the strictest direct detection bound LZ-2022, the survivable parameter space are obtained. It then turns out that the constraints on parameters $M_1$ ($\approx m_{\tilde{\chi}^0_1}$), $M_2$ ($\approx m_{\tilde{\chi}^{0/\pm}_{\rm{wino-like}}}$), $m_{\tilde{\mu}_{LR}}$ ($\approx m_{\tilde{\mu}_1}$), $\mu,\mu_T$ and $\tan\beta$ are very strong, and they are shown explicitly in the $Numerical~analysis$ section. The mass of bino-like LSP is expected to be no more than $450~\rm{GeV}$, and the ratio $\mu_T$ to $M_1$ should be within the region $[1.175,1.225]$ for the relic density of bino-like LSP to be $0.12\pm0.001$. The parameter $M_2$ within the range [700,1150] $\rm{GeV}$ are excluded, and $m_{\tilde{\mu}_{LR}}$ ($\mu$) is required to be larger than 600 (750) $\rm{GeV}$. A $\tan \beta$ exceeding 30 is also necessary for the parameter space not being ruled out.

Notably, the viable parameter space we have identified is quite different from the MSSM bino LSP scenario, where the bino LSP needs to coannihilate with winos or sleptons. A large portion of the yet unexplored regions in this parameter space can be probed in the near future, thanks to the unprecedented sensitivity of upcoming dark matter detectors, such as the Xenon-nT projection in 20 t y.\\

\noindent{\bf Acknowledgements:}
This work is supported in part by National Natural Science Foundation of China (NNSFC) under Grant No.12075074, No.12235008, No.A2023201041, No.12347101, No.12175025 and No.11705045, and by the Chongqing Graduate Research and Innovation Foundation under Grant No.ydstd1912.

\appendix
\begin{center}
\Large{{\bf Appendix}}
\end{center}
\vspace{-8mm}

\section{The Tadpole equations} \label{APPENDIX A}

The tree-level scalar potential in the TNMSSM:
\begin{eqnarray}
V&=&\frac{1}{32} \left( {g_1}^2+ {g_2}^2\right) \left(- {v_d}^2+2  {v_T}^2-2  {v_{\bar{T}}}^2+ {v_u}^2\right)^2 \nonumber\\
&&+\frac{ {m_{H_d}^2}  {v_d}^2}{2}+\frac{ {m_{H_u}^2}  {v_u}^2}{2}+\frac{ {m_S^2}  {v_s}^2}{2}+\frac{ {m_T^2}  {v_T}^2}{2}+\frac{ {m_{\bar{T}}^2}  {v_{\bar{T}}}^2}{2}\nonumber\\
&&+2 \left(\frac{ {T_{\kappa}  }  {v_s}^3}{6 \sqrt{2}}-\frac{ {T_{\lambda}  }  {v_d}  {v_s}  {v_u}}{2 \sqrt{2}}-\frac{ {T_{\lambda_T}}  {v_s}  {v_T}  {v_{\bar{T}}}}{2 \sqrt{2}}+\frac{ {T_{\chi_d}}  {v_d}^2  {v_T}}{2 \sqrt{2}}+\frac{ {T_{\chi_u}}  {v_{\bar{T}}}  {v_u}^2}{2 \sqrt{2}}\right)\nonumber\\
&&+\frac{1}{4} \left( {v_d}^2  { \chi_d}- { \lambda_T}  {v_s}  {v_{\bar{T}}}\right)^2+\frac{1}{4} \left(-\lambda   {v_d}  {v_u}+\kappa   {v_s}^2- { \lambda_T}  {v_T}  {v_{\bar{T}}}\right)^2+\frac{1}{4} (2  {v_d}  {v_T}  { \chi_d}-\lambda   {v_s}  {v_u})^2\nonumber\\
&&+\frac{1}{4} (2  {v_{\bar{T}}}  {v_u}  { \chi_u}-\lambda   {v_d}  {v_s})^2+\frac{1}{4} \left( {v_u}^2  { \chi_u}- { \lambda_T}  {v_s}  {v_T}\right)^2.
\end{eqnarray}

The tadpole equations:
\begin{eqnarray}
\frac{\partial V}{\partial v_{d}} & = & \frac{1}{8}(g_{1}^{2}+g_{2}^{2})v_d (2 v_{\bar{T}}^{2}-2 v_{T}^{2}-v_{u}^{2}+v_{d}^{2})\nonumber \\
&& +\frac{1}{4}(v_{\bar{T}}((-2 v_{d} v_{s} \chi_{d}+v_{T} v_{u} \lambda) \lambda_{T}-2 v_{s} v_{u} \lambda \chi_{u})-v_{s}^{2} v_{u} \lambda \kappa \nonumber\\
&& +(2 v_{d}(v_{s}^{2}+v_{u}^{2}) \lambda+v_{u}(\lambda_{T} v_{T} v_{\bar{T}}-v_{s}(2(v_{\bar{T}} \chi_{u}+v_{T} \chi_{d})+v_{s} \kappa))) \lambda\nonumber\\
&&-2(\sqrt{2} v_{s} v_{u} T_{\lambda}+(v_{d}(-2(2 v_{T}^{2}+v_{d}^{2}) \chi_{d}+\lambda_{T} v_{s} v_{\bar{T}})+v_{s} v_{T} v_{u} \lambda) \chi_{d})) \nonumber \\
&&+\sqrt{2} v_{T}v_d T_{\chi_{d}}+m_{H_{d}}^{2}v_d=0,\\
\frac{\partial V}{\partial v_{u}} & = & \frac{1}{8}(g_{1}^{2}+g_{2}^{2})v_u (-2 v_{\bar{T}}^{2}+2 v_{T}^{2}-v_{d}^{2}+v_{u}^{2}) \nonumber\\
&& +\frac{1}{4}((4 v_{u}^{3}+8 v_{\bar{T}}^{2} v_{u})\chi_{u}^{2}+v_{T}(-2 v_{d} v_{s} \lambda \chi_{d}+(-2 v_{s} v_{u} \chi_{u}+v_{d} v_{\bar{T}} \lambda) \lambda_{T} ) \nonumber \\
&& + (2 (v_{d}^{2}+v_{s}^{2} ) v_{u} \lambda+v_{d} (\lambda_{T} v_{T} v_{\bar{T}}-v_{s} (2 (v_{\bar{T}} \chi_{u}+v_{T} \chi_{d} )+v_{s} \kappa ) ) ) \lambda \nonumber\\
&&  +v_{u} (-2 \lambda_{T} v_{s} v_{T} \chi_{u} )
+v_{d} (-2 \sqrt{2} v_{s} T_{\lambda} +\lambda (-2 v_{s} v_{\bar{T}} \chi_{u}-v_{s}^{2} \kappa) ) )\nonumber\\
&&+\sqrt{2} v_{\bar{T}}v_u T_{\chi_{u}} +m_{H_{u}}^{2}v_u=0, \\
\frac{\partial V}{\partial v_{s}} & = & \frac{1}{4} ( (-v_{d}^{2} v_{\bar{T}} \chi_{d}+v_{s} (2 \lambda_{T} (v_{T}^{2}+v_{\bar{T}}^{2} )-2 v_{T} v_{\bar{T}} \kappa )-v_{T} v_{u}^{2} \chi_{u} ) \lambda_{T}\nonumber\\
&&+ (-2 v_{d} v_{T} v_{u} \lambda-\lambda_{T} v_{d}^{2} v_{\bar{T}} ) \chi_{d} + (-2 v_{d} v_{\bar{T}} v_{u} \lambda-\lambda_{T} v_{T} v_{u}^{2} ) \chi_{u}\nonumber\\
&&+ (-2 v_{d} v_{s} v_{u} \lambda+4 v_{s}^{3} \kappa ) \kappa+v_{s} (-2 \lambda_{T} v_{T} v_{\bar{T}} \kappa ) \nonumber\\
&&  +2 (-v_{d} v_{u} (v_{\bar{T}} \chi_{u}+v_{s} \kappa+v_{T} \chi_{d} )+v_{s} (v_{d}^{2}+v_{u}^{2} ) \lambda ) \lambda\nonumber\\
&&+\sqrt{2} (-2 v_{d} v_{u} T_{\lambda} -2 v_{T} v_{\bar{T}} T_{\lambda_{T}} +2v_{s}^{2} T_{\kappa} ) )+m_{S}^{2}v_s=0,
\\
\frac{\partial V}{\partial v_{T}} & = & \frac{1}{4} (g_{1}^{2}+g_{2}^{2} ) v_T(-2 v_{\bar{T}}^{2}+2 v_{T}^{2}-v_{d}^{2}+v_{u}^{2} ) \nonumber\\
&& +\frac{1}{4} ( (2 \lambda_{T} v_{T} (v_{s}^{2}+v_{\bar{T}}^{2} )+v_{d} v_{\bar{T}} v_{u} \lambda-v_{s} (v_{s} v_{\bar{T}} \kappa+v_{u}^{2} \chi_{u} ) ) \lambda_{T} \nonumber \\
&& +\lambda_{T} (-v_{s}^{2} v_{\bar{T}} \kappa-v_{s} v_{u}^{2} \chi_{u} )+v_{d} (-2v_{s} v_{u} \lambda \chi_{d}+v_{u} (-2 v_{s} \chi_{d}+\lambda_{T} v_{\bar{T}} ) \lambda ) \nonumber\\
&&  -2 \sqrt{2} v_{s} v_{\bar{T}} T_{\lambda_{T}} +2 \sqrt{2} v_{d}^{2} T_{\chi_{d}}  ) +2 v_{d}^2v_T \chi_{d}^2+m_{T}^{2}v_T=0,\\
\frac{\partial V}{\partial v_{\bar{T}}} & = & \frac{1}{4} (g_{1}^{2}+g_{2}^{2} ) v_{\bar{T}} (2 v_{\bar{T}}^{2}-2 v_{T}^{2}-v_{u}^{2}+v_{d}^{2} )\nonumber \\
&& +\frac{1}{4} ( (2 \lambda_{T} (v_{s}^{2}+v_{T}^{2} ) v_{\bar{T}}+v_{d} v_{T} v_{u} \lambda-v_{s} (v_{d}^{2} \chi_{d}+v_{s} v_{T} \kappa ) ) \lambda_{T} \nonumber \\
&& -\lambda_{T} v_{s}^{2} v_{T} \kappa+v_{d} v_{u} (-2 v_{s} \chi_{u}+\lambda_{T} v_{T} ) \lambda+v_{s} (-2 (\sqrt{2} v_{T} T_{\lambda_{T}} +v_{d} v_{u} \lambda \chi_{u} )\nonumber\\
&&-\lambda_{T} v_{d}^{2} \chi_{d} )  +2\sqrt{2} v_{u}^{2} T_{\chi_{u}} )+2 v_{u}^{2}v_{\bar{T}} \chi_{u}^{2} +m_{\bar{T}}^{2}v_{\bar{T}}=0.
\end{eqnarray}

\section{The Feynman diagrams involving coannihilation} \label{APPENDIX B}
\begin{figure}[h!]
\begin{center}
\begin{minipage}[c]{0.48\textwidth}
\includegraphics[width=2.8in]{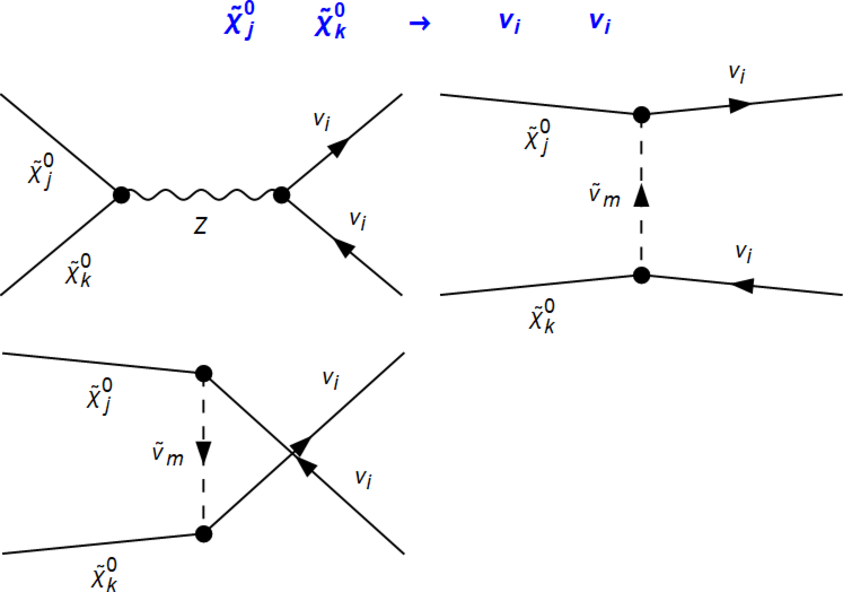}
\end{minipage}%
\begin{minipage}[c]{0.48\textwidth}
\includegraphics[width=2.8in]{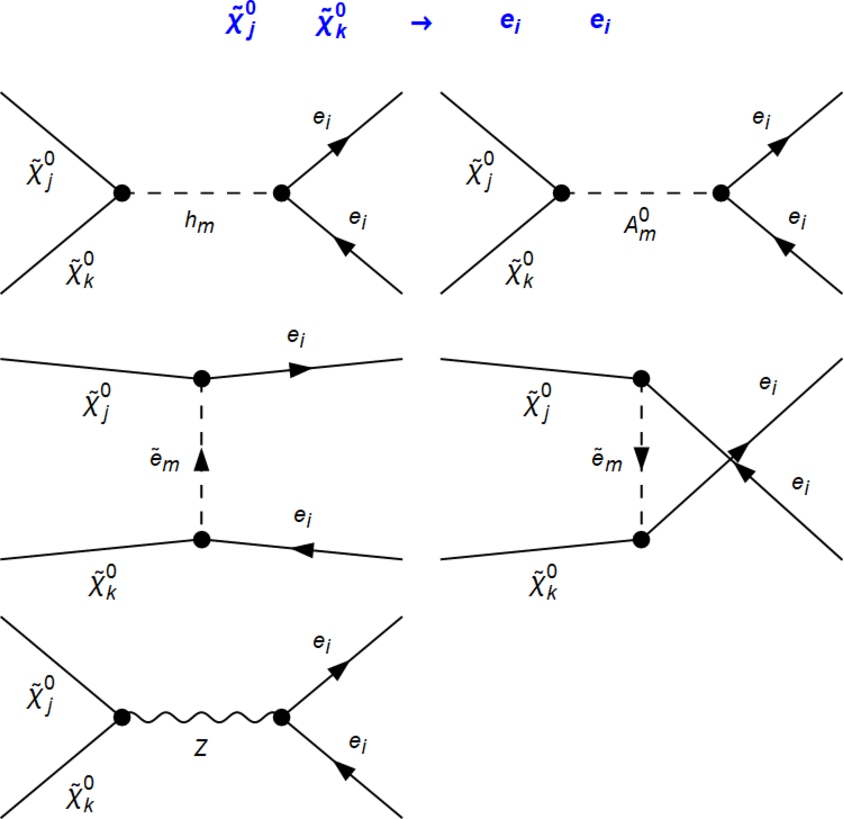}
\end{minipage}\\
\begin{minipage}[c]{0.48\textwidth}
\includegraphics[width=2.8in]{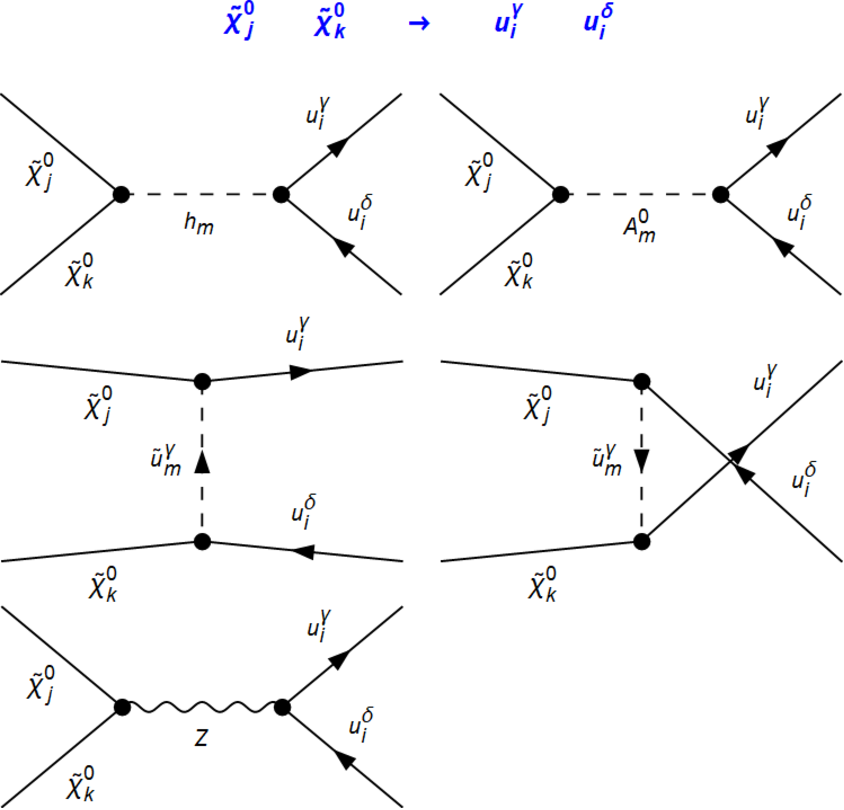}
\end{minipage}%
\begin{minipage}[c]{0.48\textwidth}
\includegraphics[width=2.8in]{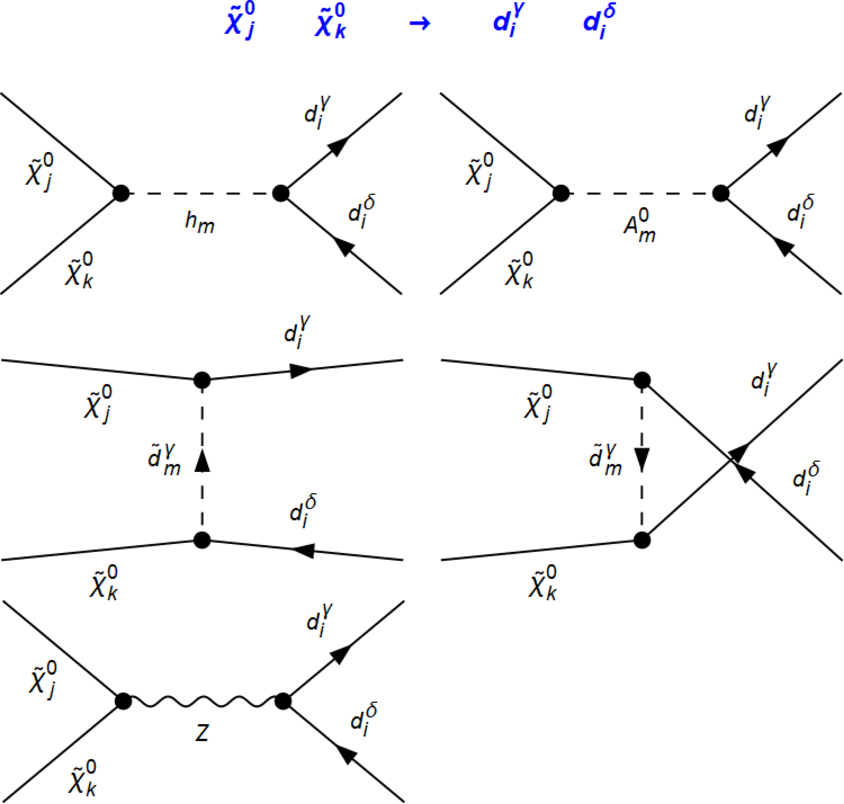}
\end{minipage}\\
\begin{minipage}[c]{0.48\textwidth}
\includegraphics[width=2.8in]{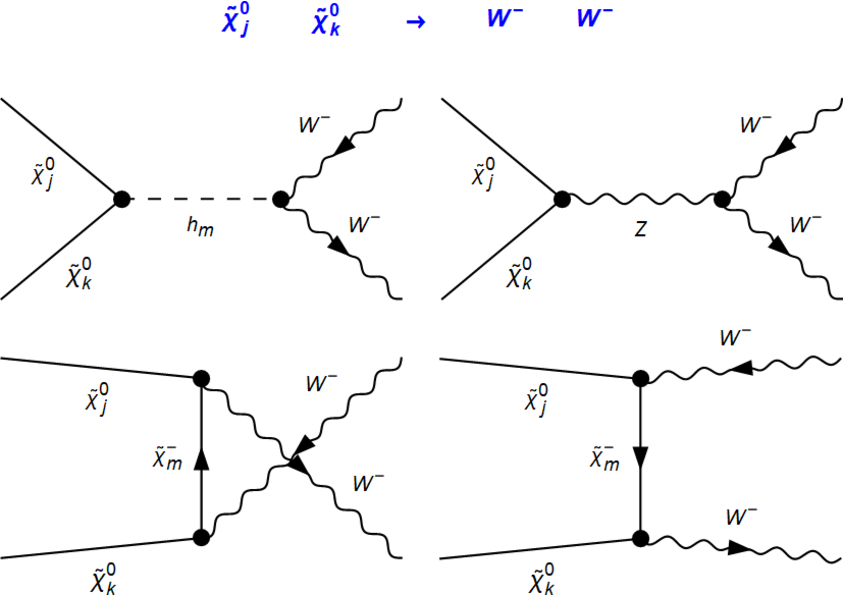}
\end{minipage}%
\begin{minipage}[c]{0.48\textwidth}
\includegraphics[width=2.8in]{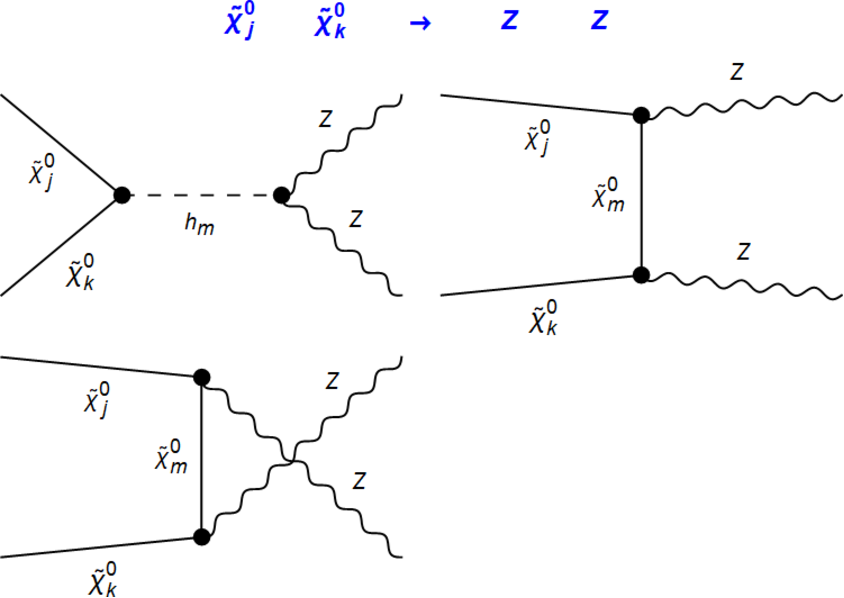}
\end{minipage}
\caption[]{\label{co1-1} Feynman diagrams for $\tilde{\chi}^0_j + \tilde{\chi}^0_k \rightarrow SM + SM$ in triplino NLSPs coannihilation scenario.}
\end{center}
\end{figure}

\begin{figure}[h!]
\begin{center}
\begin{minipage}[c]{0.48\textwidth}
\includegraphics[width=2.8in]{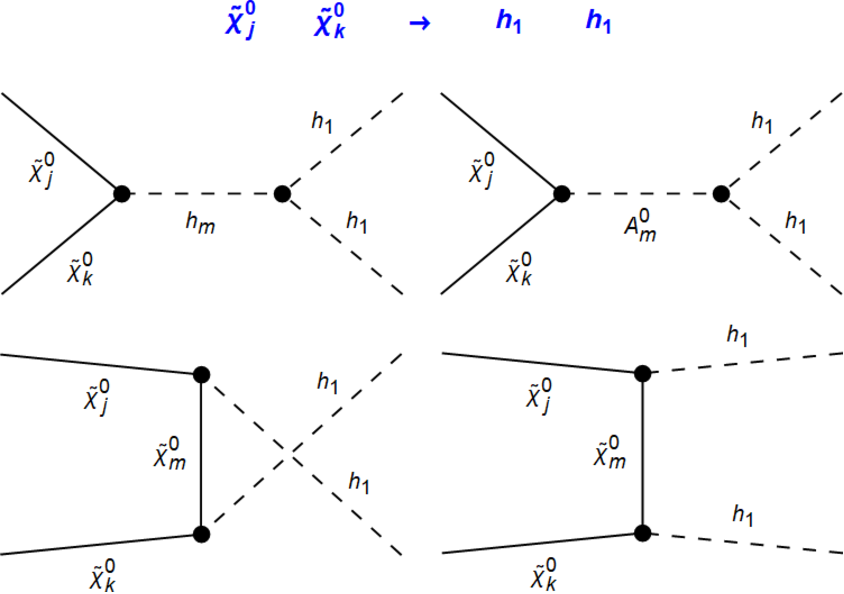}
\end{minipage}%
\begin{minipage}[c]{0.48\textwidth}
\includegraphics[width=2.8in]{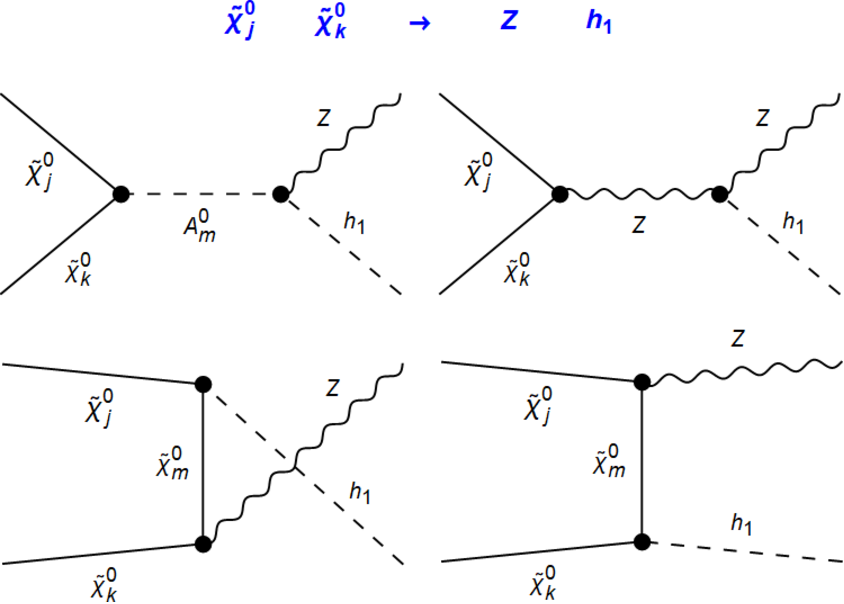}
\end{minipage}//
\begin{minipage}[c]{0.48\textwidth}
\includegraphics[width=2.8in]{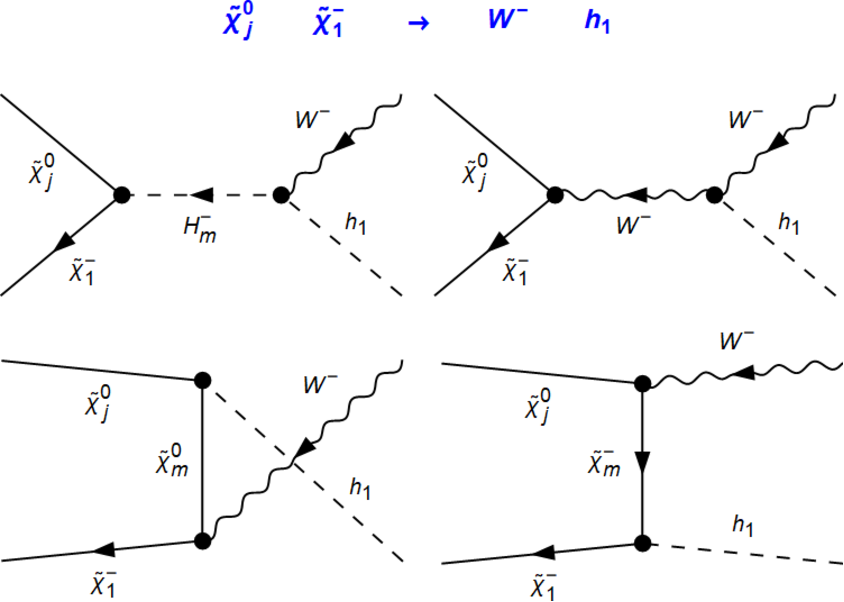}
\end{minipage}%
\begin{minipage}[c]{0.48\textwidth}
\includegraphics[width=2.8in]{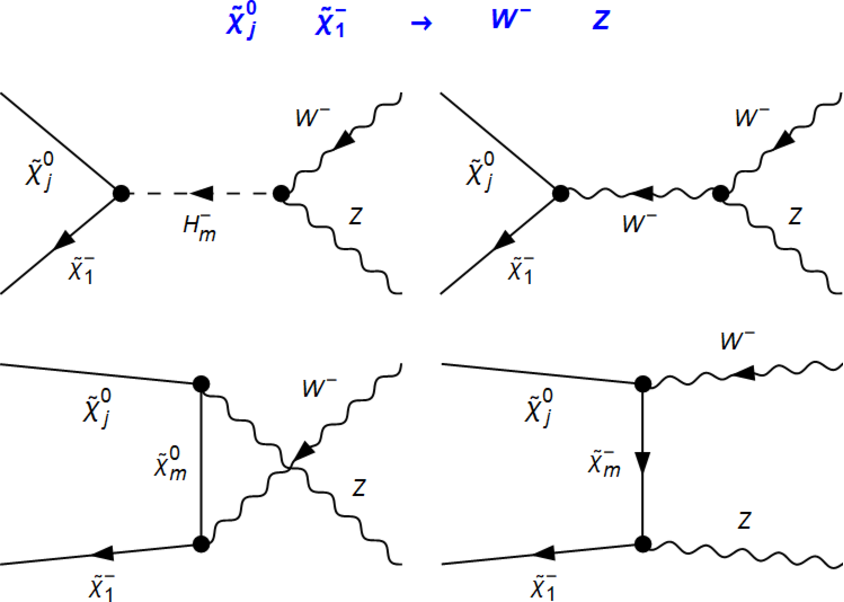}
\end{minipage}\\
\begin{minipage}[c]{0.48\textwidth}
\includegraphics[width=2.8in]{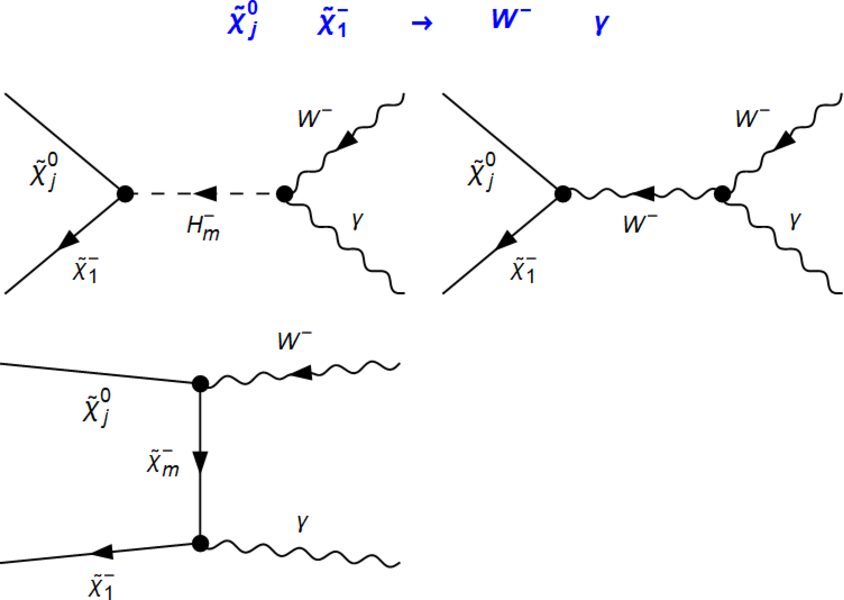}
\end{minipage}%
\begin{minipage}[c]{0.48\textwidth}
\includegraphics[width=2.8in]{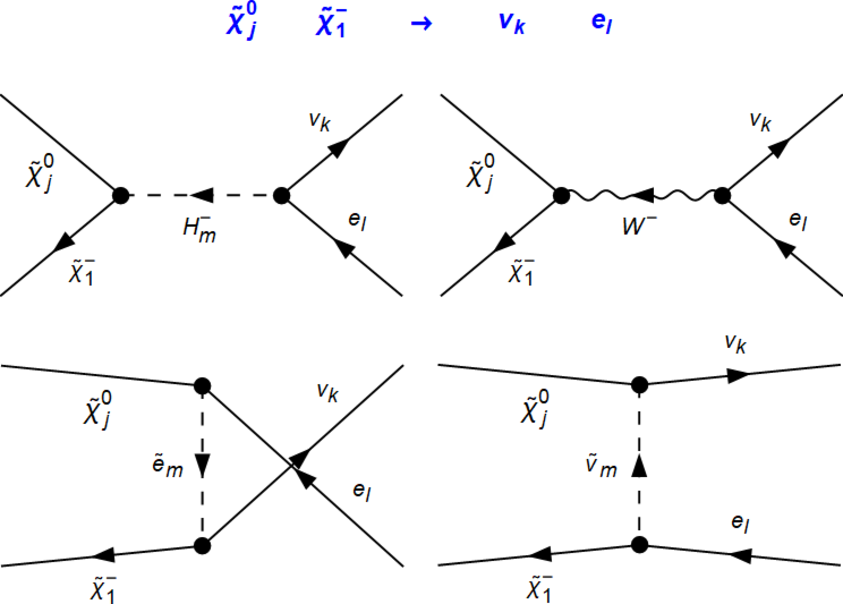}
\end{minipage}\\
\begin{minipage}[c]{0.48\textwidth}
\includegraphics[width=2.8in]{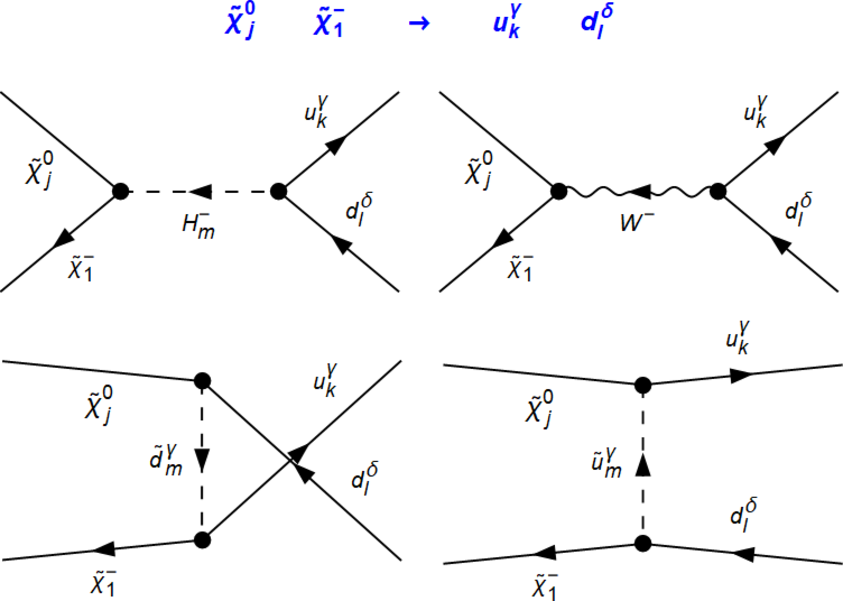}
\end{minipage}
\begin{minipage}[c]{0.48\textwidth}
\includegraphics[width=2.8in]{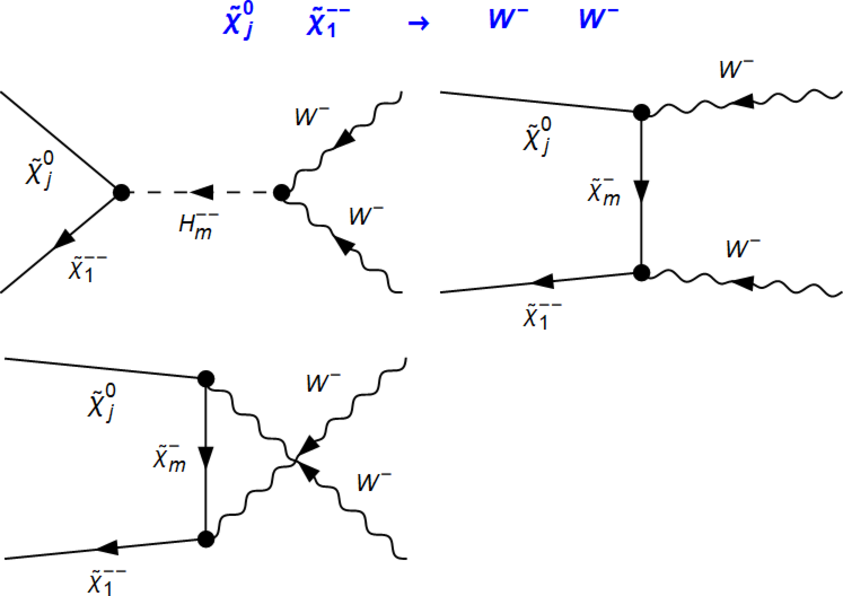}
\end{minipage}
\caption[]{\label{co2-1} Feynman diagrams for $\tilde{\chi}^0_j + \tilde{\chi}^0_k/\tilde{\chi}^-_1/\tilde{\chi}^{--}_1 \rightarrow SM + SM$ in triplino NLSPs coannihilation scenario.}
\end{center}
\end{figure}

\begin{figure}[h!]
\begin{center}
\begin{minipage}[c]{0.48\textwidth}
\includegraphics[width=2.8in]{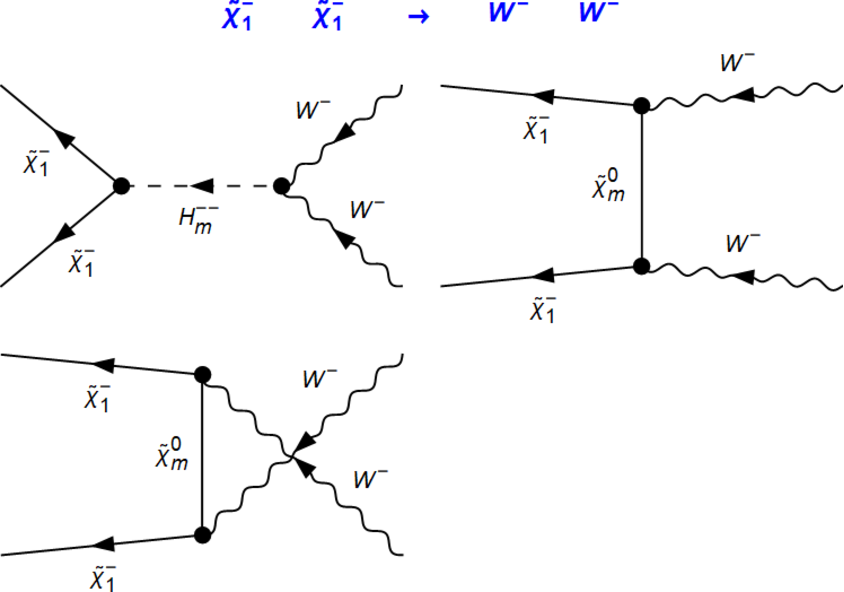}
\end{minipage}%
\begin{minipage}[c]{0.48\textwidth}
\includegraphics[width=2.8in]{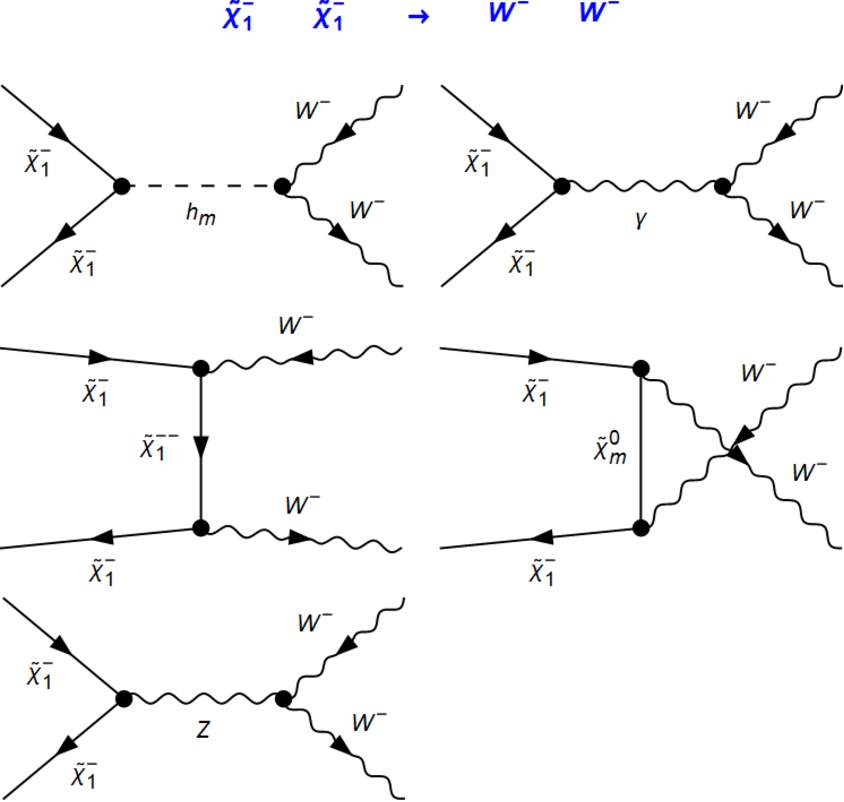}
\end{minipage}\\
\begin{minipage}[c]{0.48\textwidth}
\includegraphics[width=2.8in]{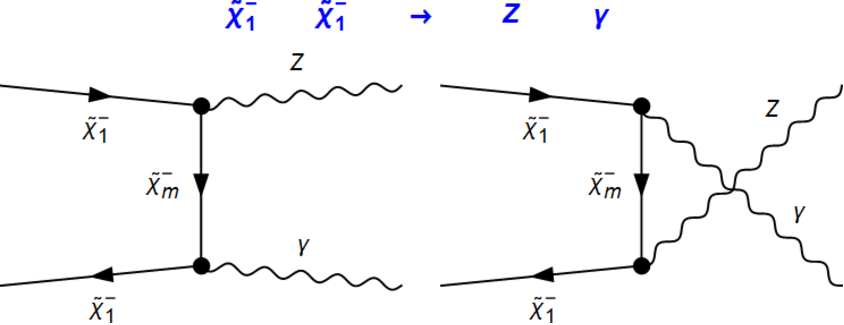}
\end{minipage}%
\begin{minipage}[c]{0.48\textwidth}
\includegraphics[width=2.8in]{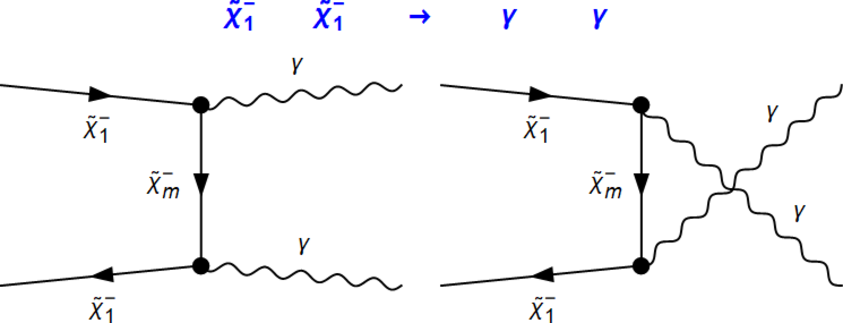}
\end{minipage}\\
\begin{minipage}[c]{0.48\textwidth}
\includegraphics[width=2.8in]{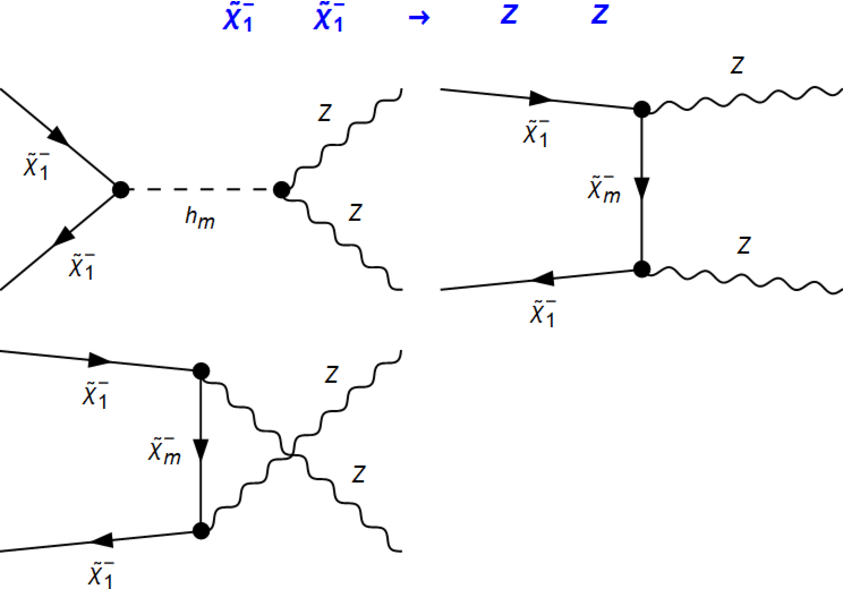}
\end{minipage}%
\begin{minipage}[c]{0.48\textwidth}
\includegraphics[width=2.8in]{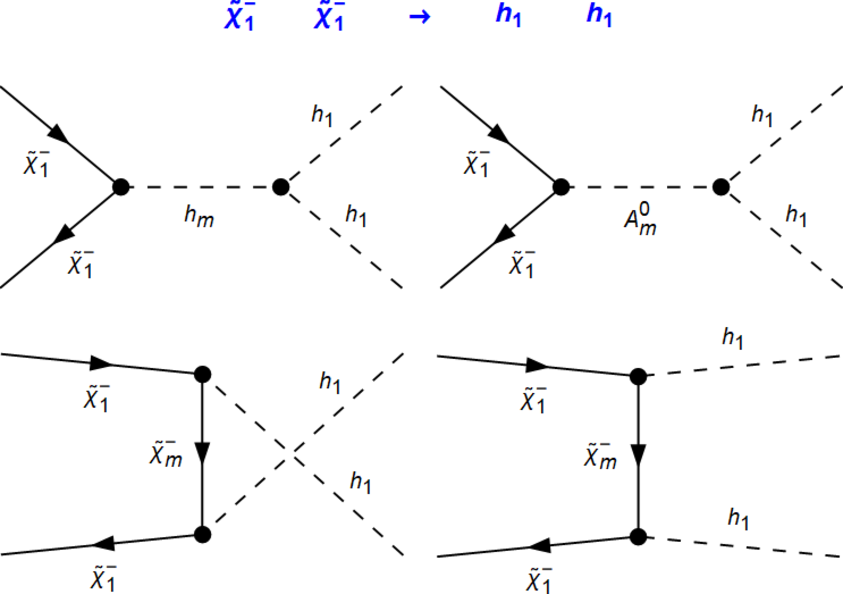}
\end{minipage}
\caption[]{\label{co3-1} Feynman diagrams for $\tilde{\chi}^-_1 + \tilde{\chi}^-_1 \rightarrow SM + SM$ in triplino NLSPs coannihilation scenario.}
\end{center}
\end{figure}

\begin{figure}[h!]
\begin{center}
\begin{minipage}[c]{0.48\textwidth}
\includegraphics[width=2.8in]{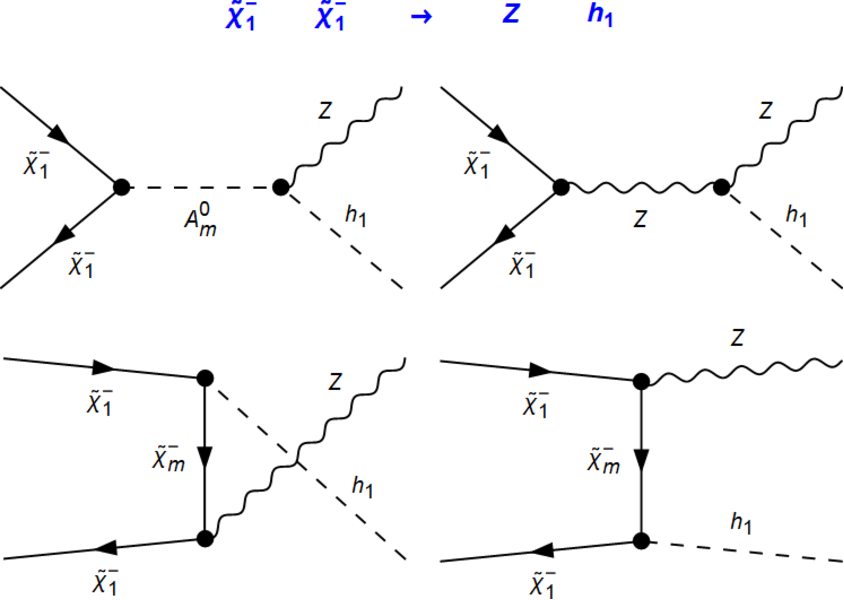}
\end{minipage}%
\begin{minipage}[c]{0.48\textwidth}
\includegraphics[width=2.8in]{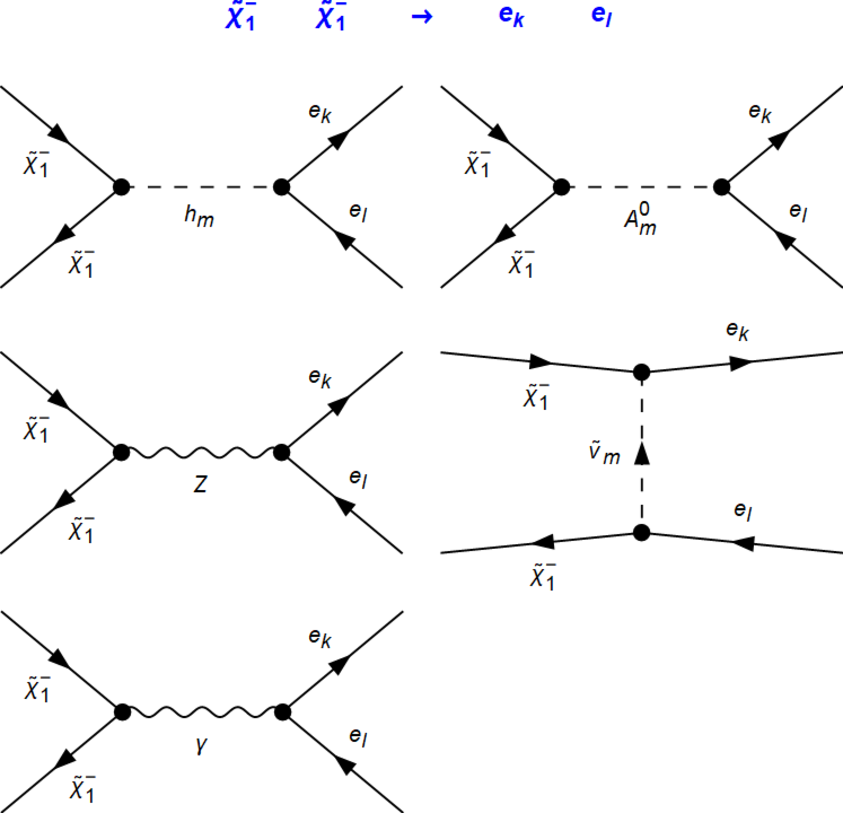}
\end{minipage}\\
\begin{minipage}[c]{0.48\textwidth}
\includegraphics[width=2.8in]{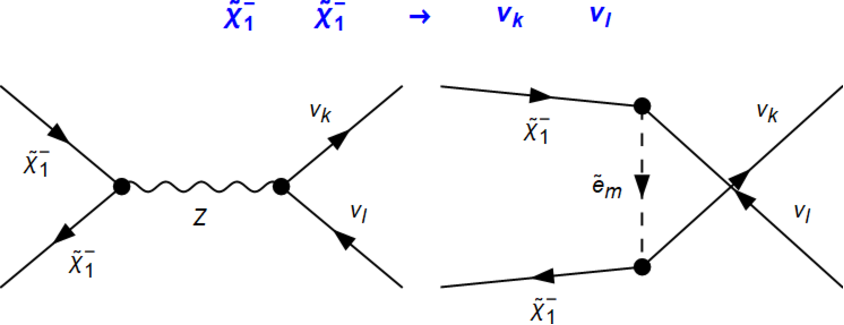}
\end{minipage}%
\begin{minipage}[c]{0.48\textwidth}
\includegraphics[width=2.8in]{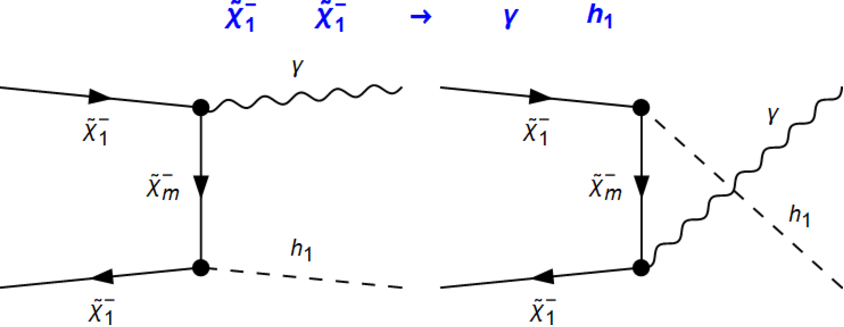}
\end{minipage}\\
\begin{minipage}[c]{0.48\textwidth}
\includegraphics[width=2.8in]{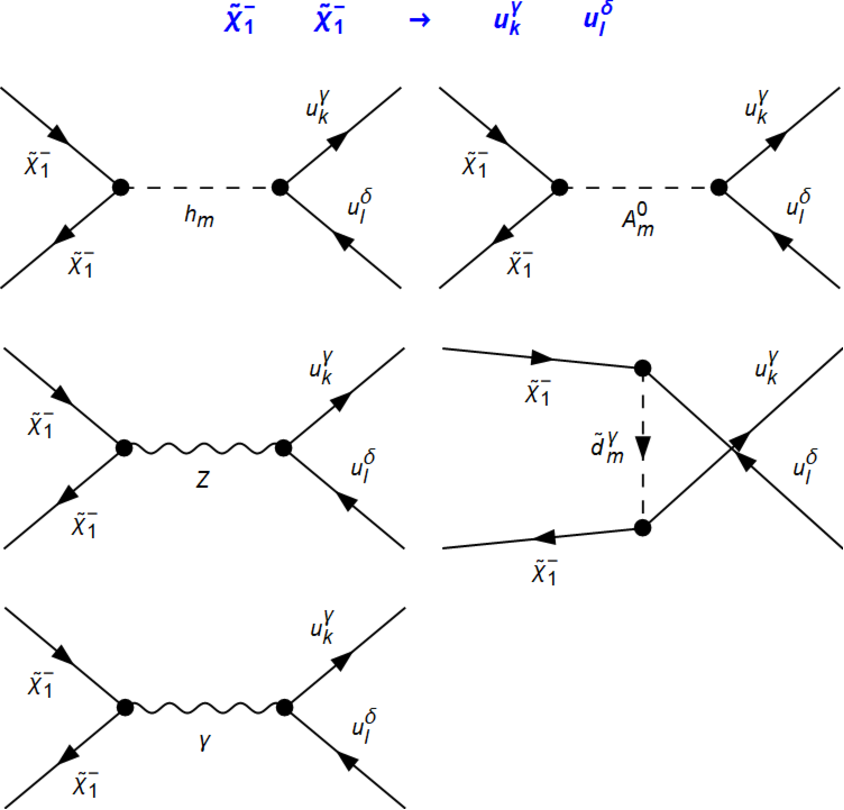}
\end{minipage}%
\begin{minipage}[c]{0.48\textwidth}
\includegraphics[width=2.8in]{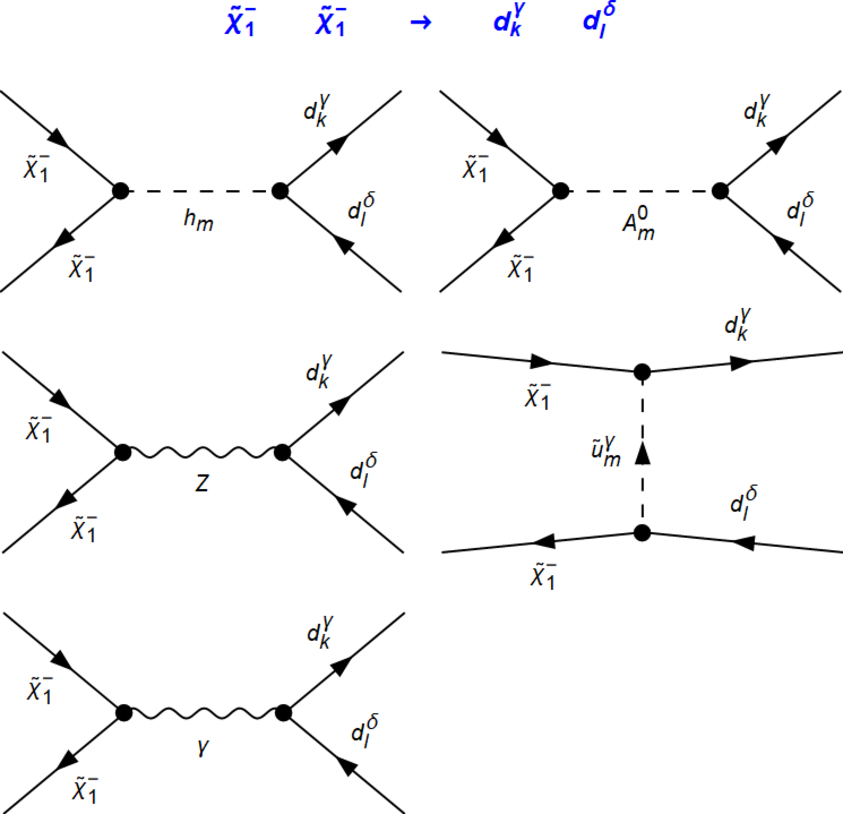}
\end{minipage}
\caption[]{\label{co3-2} Feynman diagrams for $\tilde{\chi}^-_1 + \tilde{\chi}^-_1 \rightarrow SM + SM$ in triplino NLSPs coannihilation scenario.}
\end{center}
\end{figure}

\begin{figure}[h!]
\begin{center}
\begin{minipage}[c]{0.48\textwidth}
\includegraphics[width=2.8in]{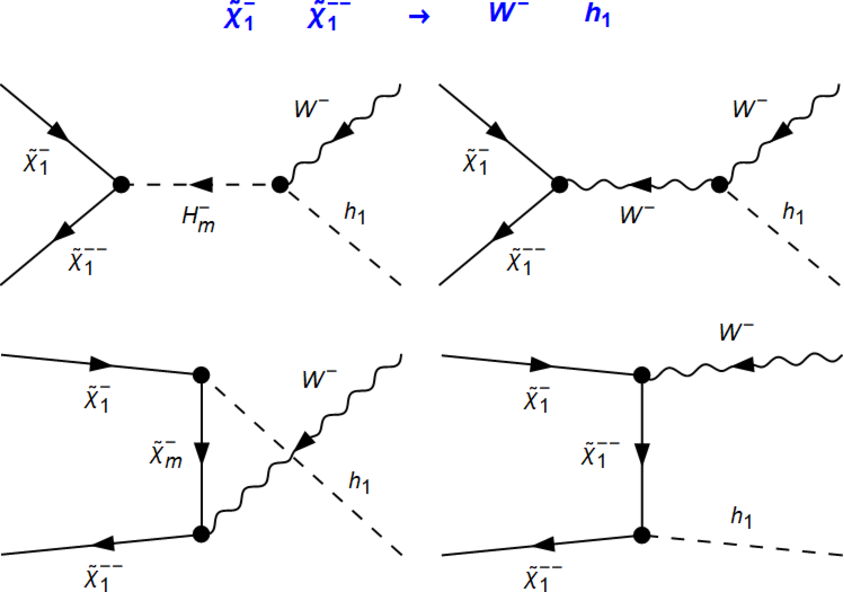}
\end{minipage}%
\begin{minipage}[c]{0.48\textwidth}
\includegraphics[width=2.8in]{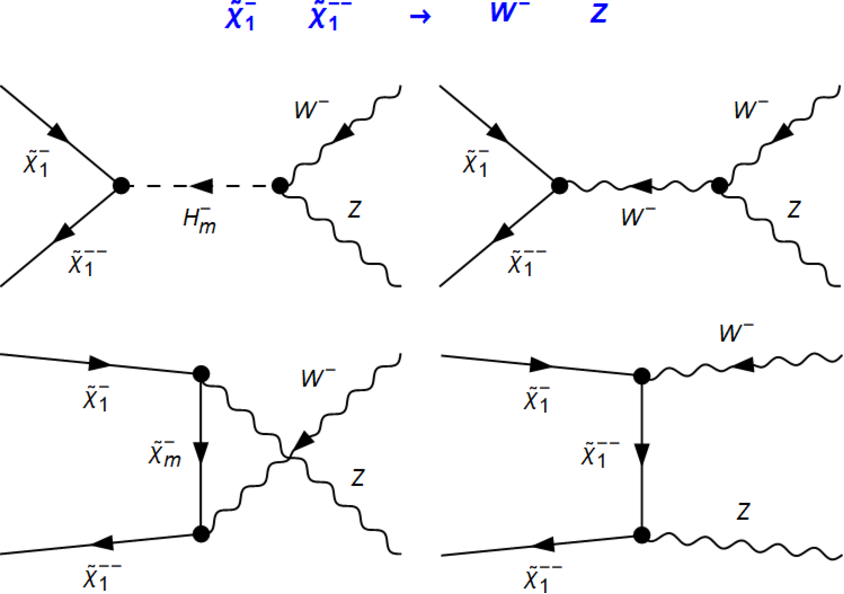}
\end{minipage}\\
\begin{minipage}[c]{0.48\textwidth}
\includegraphics[width=2.8in]{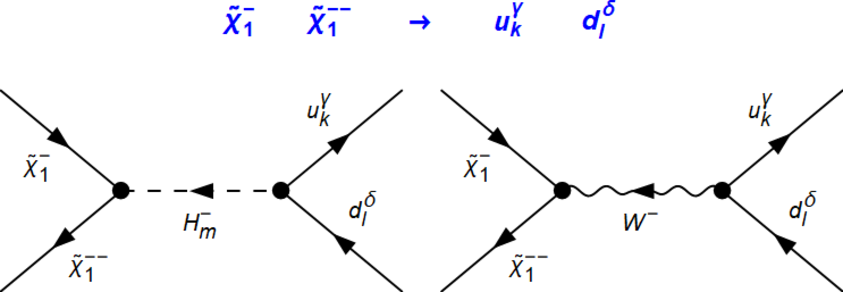}
\end{minipage}%
\begin{minipage}[c]{0.48\textwidth}
\includegraphics[width=2.8in]{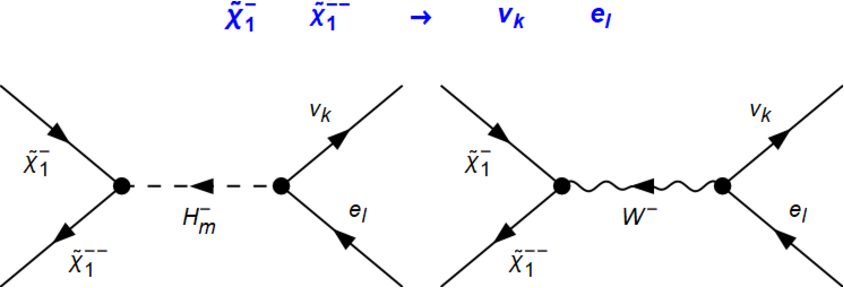}
\end{minipage}\\
\begin{minipage}[c]{0.48\textwidth}
\includegraphics[width=2.8in]{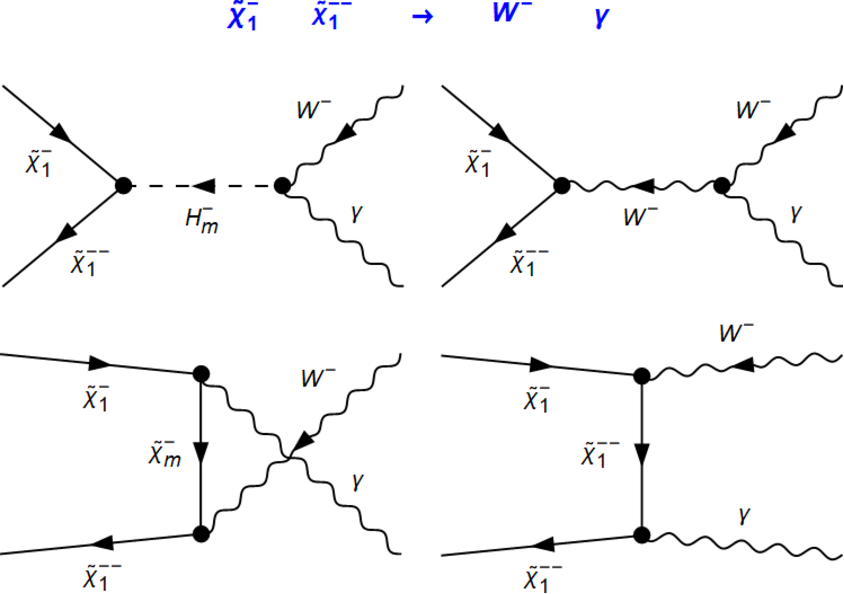}
\end{minipage}
\caption[]{\label{co4-1} Feynman diagrams for $\tilde{\chi}^-_1 + \tilde{\chi}^{--}_1 \rightarrow SM + SM$ in triplino NLSPs coannihilation scenario.}
\end{center}
\end{figure}

\begin{figure}[h!]
\begin{center}
\begin{minipage}[c]{0.48\textwidth}
\includegraphics[width=2.8in]{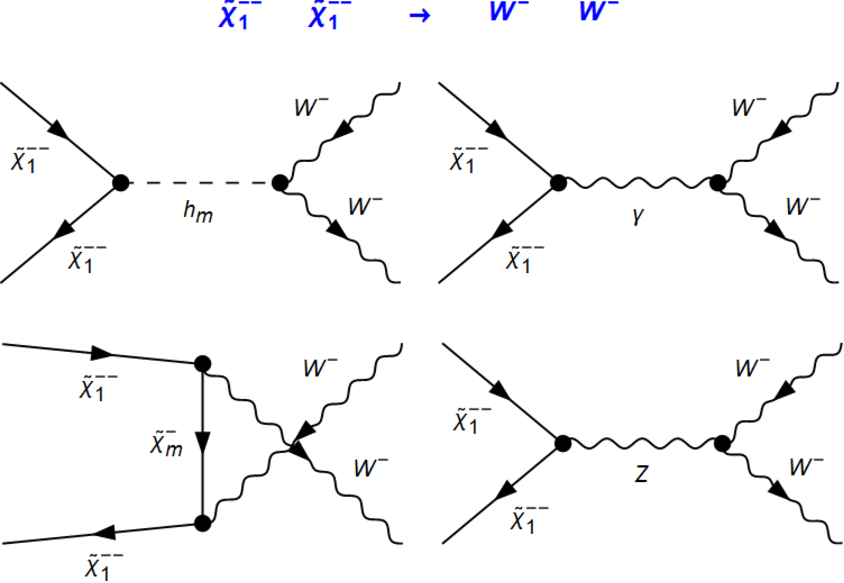}
\end{minipage}%
\begin{minipage}[c]{0.48\textwidth}
\includegraphics[width=2.8in]{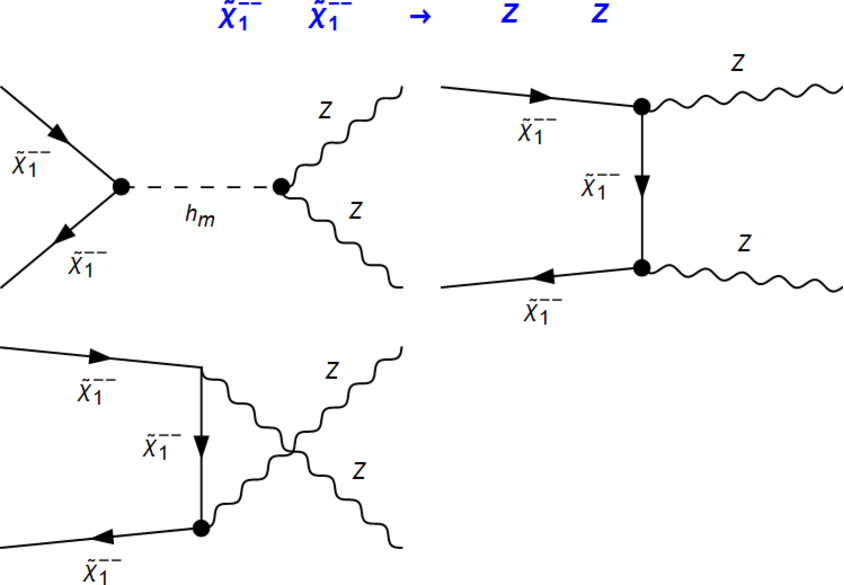}
\end{minipage}\\
\begin{minipage}[c]{0.48\textwidth}
\includegraphics[width=2.8in]{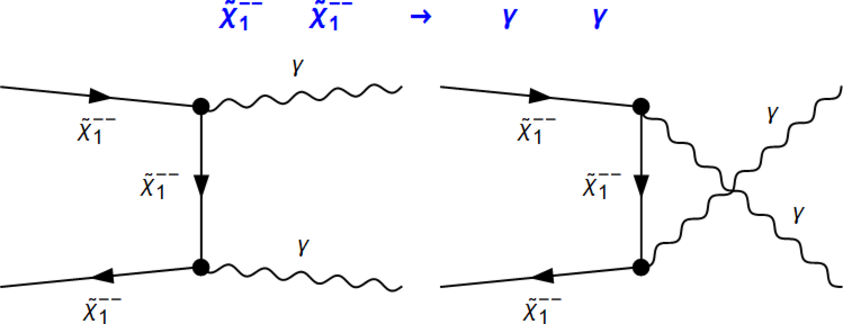}
\end{minipage}%
\begin{minipage}[c]{0.48\textwidth}
\includegraphics[width=2.8in]{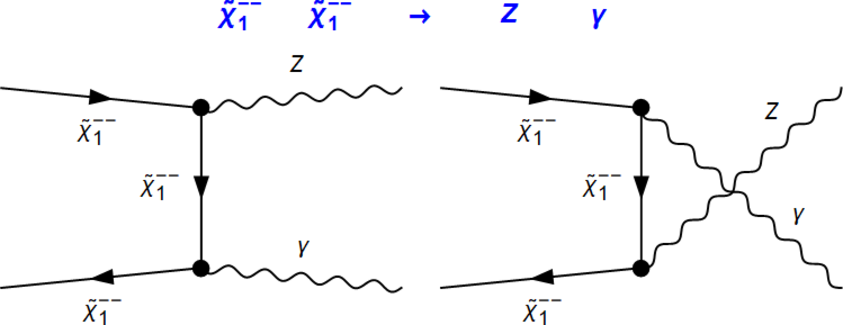}
\end{minipage}\\
\begin{minipage}[c]{0.48\textwidth}
\includegraphics[width=2.8in]{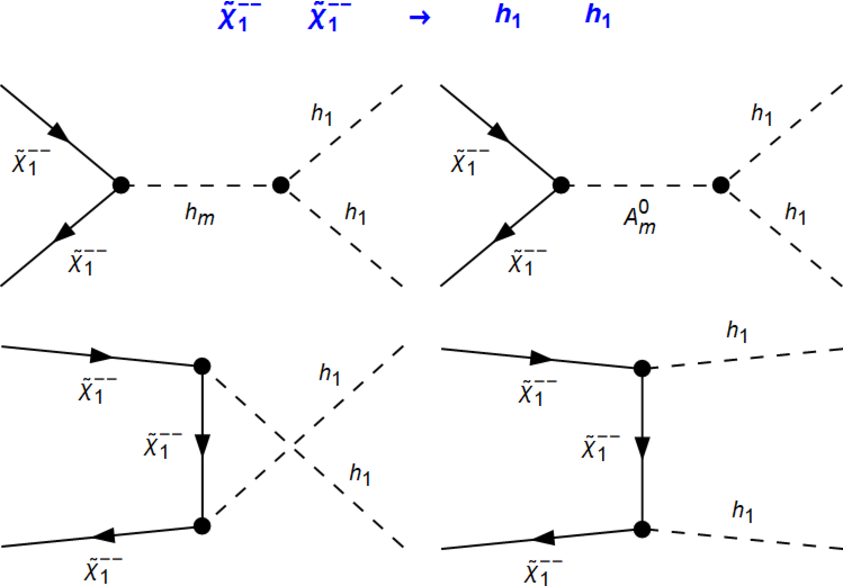}
\end{minipage}
\begin{minipage}[c]{0.48\textwidth}
\includegraphics[width=2.8in]{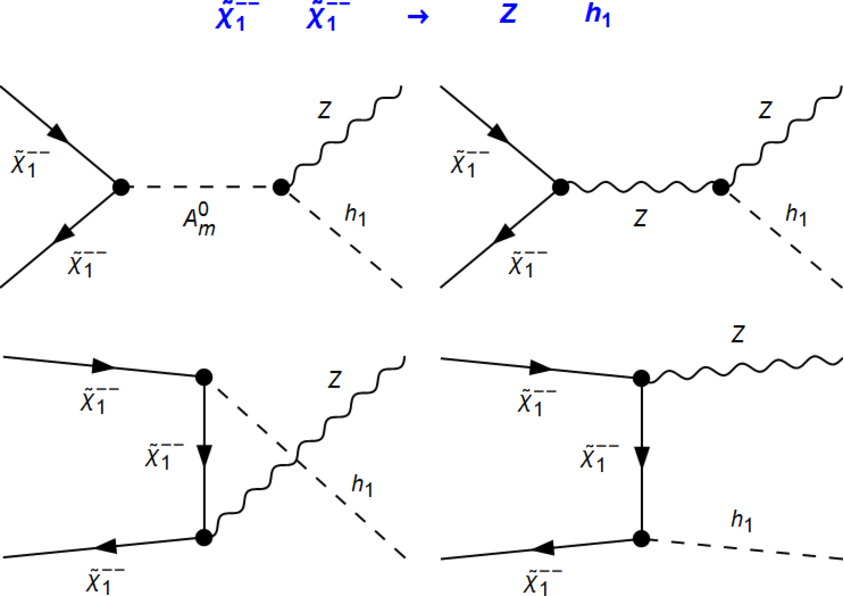}
\end{minipage}\\
\begin{minipage}[c]{0.48\textwidth}
\includegraphics[width=2.8in]{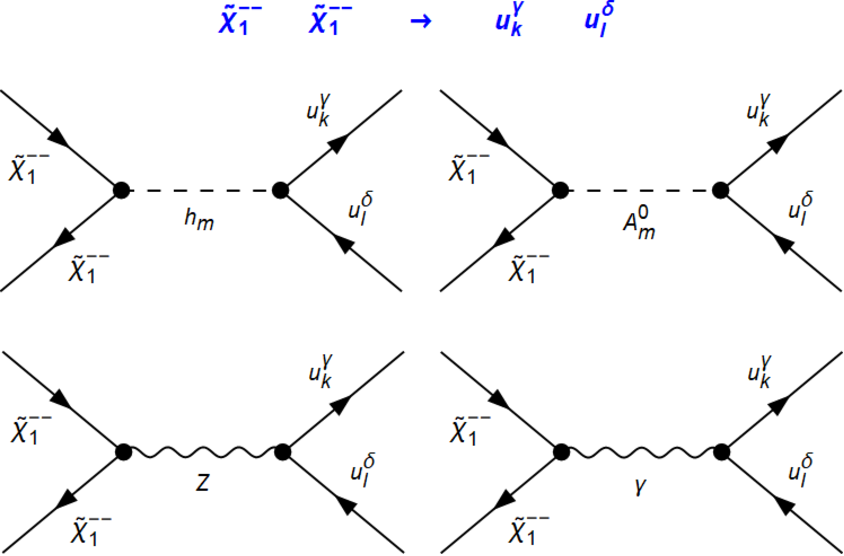}
\end{minipage}%
\begin{minipage}[c]{0.48\textwidth}
\includegraphics[width=2.8in]{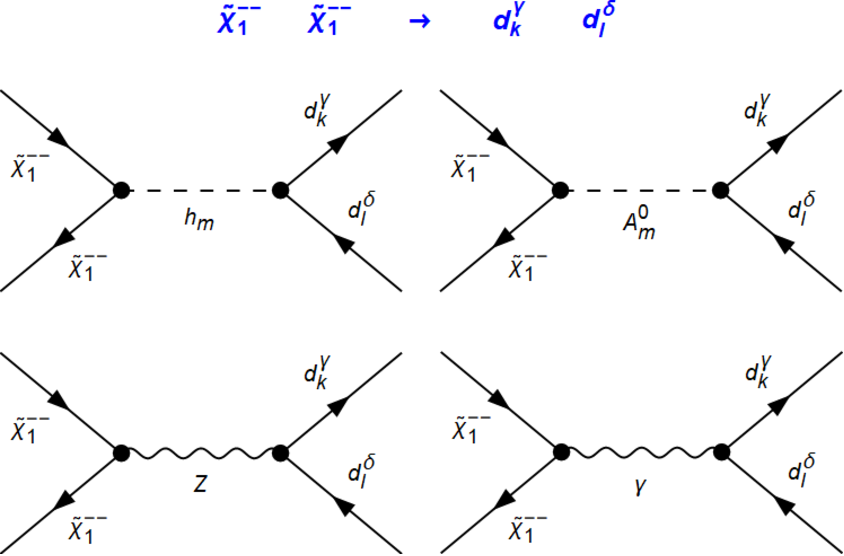}
\end{minipage}
\caption[]{\label{co5-1} Feynman diagrams for $\tilde{\chi}^{--}_1 + \tilde{\chi}^{--}_1 \rightarrow SM + SM$ in triplino NLSPs coannihilation scenario.}
\end{center}
\end{figure}

\begin{figure}[h!]
\begin{center}
\begin{minipage}[c]{0.48\textwidth}
\includegraphics[width=2.8in]{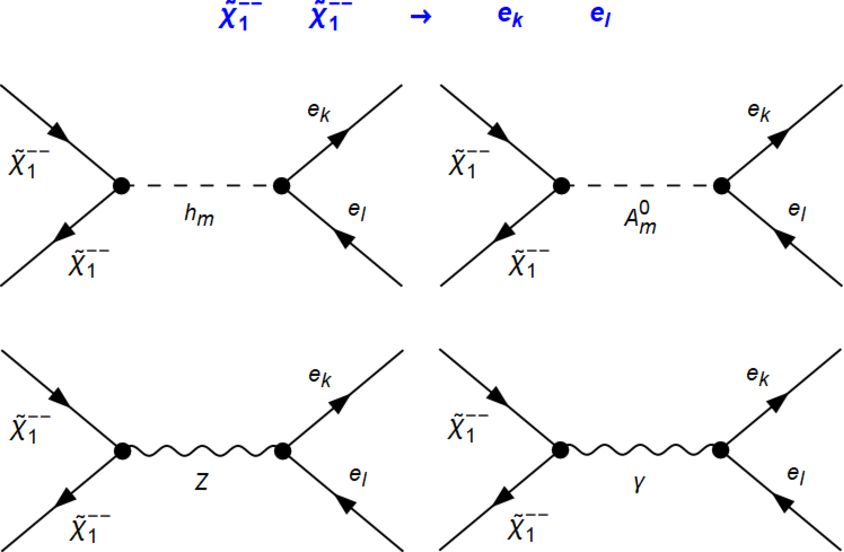}
\end{minipage}\\
\begin{minipage}[c]{0.48\textwidth}
\includegraphics[width=2.8in]{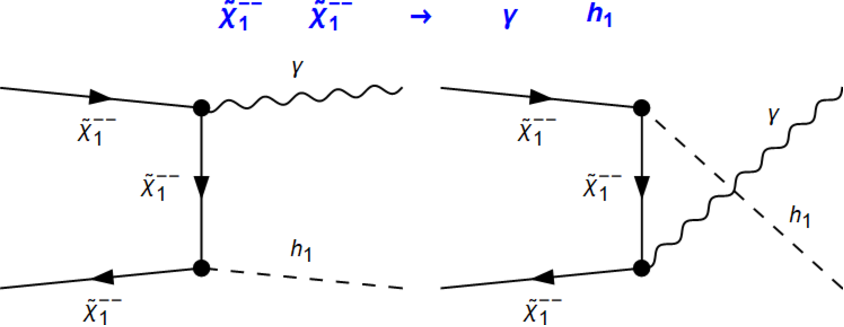}
\end{minipage}\\
\begin{minipage}[c]{0.48\textwidth}
\includegraphics[width=1.4in]{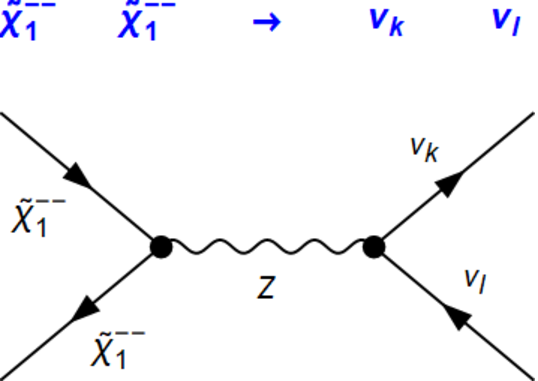}
\end{minipage}
\caption[]{\label{co5-2} Feynman diagrams for $\tilde{\chi}^{--}_1 + \tilde{\chi}^{--}_1 \rightarrow SM + SM$ in triplino NLSPs coannihilation scenario.}
\end{center}
\end{figure}

\end{document}